\documentclass[ALICE,manyauthors]{cernphprep}
\usepackage[T1]{fontenc}

\usepackage[utf8]{inputenc}
\usepackage[comma,square,numbers,sort&compress]{natbib}
\usepackage{hyperref}

\usepackage{lineno}

\usepackage{xcolor} 

\usepackage{siunitx}

\newcommand{\energy}{\ensuremath{{\sqrt{s_{\mathrm{NN}}}}}\xspace} 
\newcommand{\pt}{\ensuremath{p_{\rm T}}\xspace}
\newcommand{\mt}{\ensuremath{m_{\rm T}}\xspace}
\newcommand{\dEdx}{d$E$/d$x$\xspace}

\newcommand{\dNdy}{\ensuremath{{\rm d}N/{\rm d}y}\xspace}
\newcommand{\dndeta}{\ensuremath{\langle\text{d}N_\text{ch}/\text{d}\eta_\mathrm{lab}\rangle}\xspace}
\newcommand{\dndydpt}{\ensuremath{{\rm d}^2N/({\rm d}y {\rm d}p_{\rm T})}\xspace} 

\begin{document}%

\begin{titlepage}
\PHyear{2019}
\PHnumber{246}      
\PHdate{28 October}  
%

\title{Production of (anti-)$^3$He and (anti-)$^3$H  in p--Pb collisions
\\ at $\sqrt{s_{\rm{NN}}}$ = 5.02 TeV}
\ShortTitle{Production of (anti-)$^3$He and (anti-)$^3$H in p--Pb collisions at \energy~= 5.02 TeV}   

\Collaboration{ALICE Collaboration\thanks{See Appendix~\ref{app:collab} for the list of collaboration members}}
\ShortAuthor{ALICE Collaboration} 

\begin{abstract}
The transverse momentum (\pt) differential yields of (anti-)$^3$He and (anti-)$^3$H measured in p--Pb collisions at \energy~=~5.02~TeV with ALICE at the Large Hadron Collider (LHC) are presented. 
The ratios of the \pt-integrated yields of (anti-)$^3$He and (anti-)$^3$H to the proton yields are reported, as well as the \pt dependence of the coalescence parameters $B_3$ for (anti-)$^3$He and \mbox{(anti-)}$^3$H.
For (anti-)$^3$He, the results obtained in four classes of the mean charged-particle multiplicity density are also discussed. 
These results are compared to predictions from a canonical statistical hadronization model and coalescence approaches. An upper limit on the total yield of $^4\overline{\mathrm{He}}$ is determined.
\end{abstract}
\end{titlepage}
\setcounter{page}{2}


\section{Introduction}

In ultra-relativistic nuclear collisions, midrapidity production yields of ordinary hadrons, i.e., mesons and baryons, can be described within the Statistical Hadronization Model (SHM), for which the temperature and the baryo-chemical potential are the parameters regulating hadron production~\cite{Andronic:2017pug, BraunMunzinger:2003zd}. In this model, hadrons are produced from an expanding medium in local thermodynamic equilibrium. Their abundances are fixed when the rate of inelastic collisions becomes negligible. This chemical freeze-out is associated with a characteristic temperature which is found to be $T_\textrm{chem} \approx 156$~MeV in Pb--Pb collisions at the Large Hadron Collider (LHC) \cite{Andronic:2017pug}.
The yields of hadrons in central Pb--Pb collisions are reproduced by this approach \cite{BraunMunzinger:2003zd} within uncertainties.
Elastic and quasi elastic scattering might still occur among hadrons during the further evolution of the system. The transverse momentum distributions can be modified until also the elastic interactions cease at the kinetic freeze-out.
At LHC energies, baryon number transport from the initial nuclei at beam rapidity to midrapidity is completely negligible. This implies that particles and their corresponding antiparticles are produced in approximately equal amounts which is accounted for by a vanishing baryo-chemical potential $\mu_\mathrm{B}$.

Light (anti-)nuclei are composite objects of (anti-)baryons with radii that are substantially larger than those of ordinary hadrons and their sizes reach a significant fraction of the volume of the expanding medium. Their production yields can also be described within the SHM. This may be surprising as the separation energy of nucleons is much smaller than the system temperature, thus raising the question of how nuclei can survive during the hadronic phase.
Alternative approaches were developed that are able to describe production yields of light nuclei via the coalescence of protons and neutrons which are close by in phase space at kinetic freeze-out \cite{Kapusta:1980zz,Scheibl:1998tk}.
In this simplified approach, the invariant yield of nuclei with mass number $A$, $E_{\mathrm{A}}(\mathrm{d}^{3}N_{\mathrm{A}}/\mathrm{d}p_{\mathrm{A}}^{3})$, is related to that of nucleons via
\begin{equation}
E_{\mathrm{A}}\frac{\mathrm{d}^{3}N_{\mathrm{A}}}{\mathrm{d}p_{\mathrm{A}}^{3}} = B_{\mathrm{A}}{\left(E_{\mathrm{p}}\frac{\mathrm{d}^{3}N_{\mathrm{p}}}{\mathrm{d}p_{\mathrm{p}}^{3}}\right)^{A}} \Bigg|_{\vec{p}_{\mathrm{p}} = \vec{p}_{\mathrm{A}}/{A}},
\label{eq:BA}
\end{equation}
where $E_{\mathrm{p}}(\mathrm{d}^{3}N_{\mathrm{p}}/\mathrm{d}p_{\mathrm{p}}^{3})$ is the invariant yield of protons, which is expected to be identical to that of neutrons at midrapidity and LHC energies \cite{Bellini:2018epz}. Here, the coalescence probability is given by the parameter $B_\mathrm{A}$. Both the SHM and the coalescence approach result in similar predictions, as demonstrated for the production of deuterons~\cite{BraunMunzinger:1994iq, Steinheimer:2012tb, Vovchenko:2018fiy}. A review can be found in Ref.~\cite{Braun-Munzinger:2018hat}.

However, recent studies \cite{Bellini:2018epz} have shown a sizable difference for the $B_\mathrm{A}$ parameter as a function of the size of the particle emitting source between predictions by the SHM with kinetic freeze-out conditions from a simple hydrodynamical model and the coalescence model. Here, information from Hanbury Brown--Twiss (HBT) correlations is used to determine the source size.
This effect is more pronounced for (hyper)nuclei with larger radii. Thus, the ideal benchmark would be to study the production of hypertriton ($^3_\Lambda$H) as a function of the mean charged-particle multiplicity density, which is not yet possible due to the size of the data sets available.
Thus, the difference between the production of $^3$He and $^3$H was studied, which is expected to offer similar insight on the comparison of the SHM and the coalescence approaches, especially for smaller collision systems \cite{Mrowczynski:2016xqm,Bellini:2018epz}. 
The production yields for (anti-)$^3$He in pp and Pb--Pb \cite{Acharya:2017fvb,Adam:2015vda} collisions measured by ALICE do not cover completely the evolution from small to large source sizes. To bridge this gap, measurements in p--Pb collisions are needed which cover the intermediate source sizes.

In a broader context, the measurement of the production of nuclei in pp and p--Pb collisions contributes significantly and decisively to indirect searches for segregated primordial antimatter and dark matter via satellite-borne instruments, such as AMS-02 \cite{Kounine:2012ega}. These experiments search for an excess in the measured production of anti-nuclei above the background stemming from pp and p--A collisions in the interstellar medium. This background is predicted by calculations~\cite{Blum:2017qnn} that use measurements of the production of anti-nuclei in accelerator experiments as a key ingredient.

This paper reports on the transverse momentum differential yields of the isospin partners (anti-)$^3$He and (anti-)$^3$H in p--Pb collisions at \energy~=~5.02~TeV in the rapidity range $-1 \le y_\mathrm{cms} < 0$. In case of \mbox{(anti-)}$^3$He, the multiplicity dependence of the \pt-differential and integrated yields is also presented. An upper limit on the production yield of $^4\overline{\mathrm{He}}$ is given.
The paper is organized as follows. The experimental setup and data sample are described in Section~2. Section~3 summarizes the data analysis, while the techniques to evaluate the systematic uncertainties are presented in Section~4. The results are discussed in Section~5 and conclusions are given in Section~6. 
 

\section{Data sample and experimental apparatus}

The results presented in this paper were obtained by analyzing the data sample of p--Pb collisions at a center-of-mass energy per nucleon--nucleon pair \energy~=~5.02~TeV collected in 2016.

ALICE is a general--purpose detector system at the LHC designed to investigate high-energy heavy-ion collisions. The excellent tracking and particle identification (PID) capabilities over a broad momentum range and the low material budget make this detector ideally suited for measurements of light \mbox{(anti)}nuclei production. The characteristics of the ALICE detectors are described in detail in Refs.~\cite{Aamodt:2008zz,Abelev:2014ffa}. 

In the ALICE coordinate system, the nominal interaction point is at the origin of a right-handed Cartesian coordinate system. The \textit{z} axis
corresponds to the beam line, the \textit{x} axis points to the center of the accelerator, and the \textit{y} axis points upward.
The beam configuration was chosen such that the protons travel toward the negative \textit{z} direction and Pb nuclei travel in the positive direction of the ALICE reference frame.

 The Inner Tracking System (ITS) \cite{Aamodt:2010aa}, the Time Projection Chamber (TPC) \cite{Alme:2010ke} and the Time-of-Flight detector (TOF) \cite{Akindinov:2013tea} are the main detectors used for track reconstruction and particle identification in these analyses. 
 They are located in the central barrel within a large solenoidal magnet, which provides a homogeneous field of $\textit{B}$ = 0.5~T parallel to the beam line.

 The ITS consists of six cylindrical layers of silicon detectors, concentric and coaxial to the beam pipe, with a minimum pseudorapidity coverage $|\eta_\mathrm{lab}| < 0.9$ calculated for the nominal interaction region. Three different technologies are used for this detector: The two innermost layers consist of Silicon Pixel Detectors (SPD), the two central layers of Silicon Drift Detectors (SDD) and the two outermost layers of double-sided Silicon Strip Detectors (SSD). The radial positions of the detectors range from 3.9~cm up to 43~cm from the interaction region.
 The ITS is used in the track reconstruction and helps to improve the \pt resolution of tracks by providing high-resolution tracking points close to the beam line. Thanks to this information, the distance of closest approach (DCA) of a track to the primary vertex can be measured with a resolution below \SI{75}{\micro\metre} for tracks with \mbox{$\pt > 1$ GeV/$c$} \cite{Aamodt:2010aa,Acharya:2018qsh}.

The TPC is the main tracking device in the ALICE central barrel with a pseudorapidity coverage $|\eta_\mathrm{lab}| < 0.9$. It is used for track reconstruction and for particle identification via the measurement of the specific ionization energy loss of charged particles (\dEdx) in the TPC gas.
The TPC is cylindrical in shape, coaxial with the beam pipe, with an active gas volume ranging from 85~to~250 cm in the radial direction, and a length of 500~cm in the beam direction.
The gas mixture used, $90\%\ \mathrm{Ar}$ and $10\%\ \mathrm{CO}_{2}$ at atmospheric pressure, is characterized by low diffusion and low $\textit{Z}$. These requirements are essential to guarantee the highest possible data acquisition rate, the excellent transverse momentum resolution (ranging from about 1$\%$ at 1 GeV/$\textit{c}$ to about 3$\%$ at 10 GeV/$\textit{c}$), and the high \dEdx resolution, which is approximately 5.5$\%$ for minimum ionizing particles crossing the full detector \cite{Abelev:2014ffa}.

The TOF detector is made of Multigap Resistive Plate Chambers (MRPC), with a pseudorapidity coverage $|\eta_\mathrm{lab}| < 0.9$
\cite{Akindinov:2013tea}.
This detector is arranged in a modular structure with 18 blocks in azimuthal angle matching the TPC sectors and is used for particle identification by measuring the time of flight of charged particles.
The collision time is provided on an event-by-event basis by the TOF detector itself or by the T0 detector~\cite{Cortese:781854}. The latter consists of two arrays of Cherenkov counters (T0C and T0A) positioned around the beam pipe, on both sides of the nominal interaction point. A weighted average is performed when both detectors have measured the start time \cite{Adam:2016ilk}. The total time resolution for the analyzed data sample is $\approx 80$ ps.

The last detector used for this analysis is the V0, which consists of two scintillator hodoscopes (V0C and V0A)  \cite{Abbas:2013taa}, covering the pseudorapidity regions $-3.7 < \eta_\mathrm{lab} < -1.7$ and $2.8 < \eta_\mathrm{lab} < 5.1$. It is used to define the minimum-bias trigger, which requires a coincident signal in V0A and V0C to reduce the contamination from single--diffractive and asymmetric electromagnetic interactions. In addition, the V0A signal is proportional to the mean charged-particle multiplicity density in the direction of the Pb beam.
The minimum-bias data sample is divided into four multiplicity classes defined as percentiles of the V0A signal. These are summarized in \autoref{Table:Multiplicity}, where the corresponding mean charged-particle multiplicity densities at midrapidity $\dndeta_{|\eta_\mathrm{lab}| < 0.5}$
are also listed. These values and their uncertainties are taken from Ref.~\cite{Abelev:2013haa}.

\begin{table}[h]
\begin{center}
\centering
\caption{Summary of the V0A multiplicity classes and their corresponding mean charged-particle multiplicity densities at midrapidity. The values and their uncertainties are taken from Ref.~\cite{Abelev:2013haa}.}
\label{Table:Multiplicity}
\begin{tabular}{lc}
\hline
V0A Classes & $\dndeta_{|\eta_\mathrm{lab}| < 0.5}$\\
\hline
0--10\% & $40.6 \pm 0.9$\\
10--20\% & $30.5 \pm 0.7$\\
20--40\% & $23.2 \pm 0.4$\\
40--100\% & $10.1 \pm 0.2$\\
\hline
\end{tabular}
\end{center}
\end{table}
\section{Data analysis}
\label{Sec:DataAnalysis}

In this section, the analysis technique is described. In particular, the criteria used for the event and track selection, the signal extraction techniques used for $^{3}$H and $^{3}$He, the corrections based on Monte Carlo (MC) simulations, and the evaluation of the systematic uncertainties are illustrated and discussed.

The reconstruction efficiencies of (anti-)$^3$H and (anti-)$^3$He, the estimate of the contribution of secondary nuclei produced by spallation in the detector material, and the subtraction of the feed-down from the weak decay of hypertriton are obtained using Monte Carlo simulations.
Nuclei and antinuclei were generated with a flat distribution in transverse momentum and rapidity within
$0 \leq \pt \le 8$ GeV/$c$ and $-1 \leq y_\mathrm{cms} \leq 1$. Ten deuterons, $^3$H, $^3$He, and $^4$He as well as their antinuclei were injected into each p--Pb collision simulated with the EPOS-LHC event generator \cite{Pierog:2013ria}. In addition, 20 hypertritons and antihypertritons were injected per event. For particle propagation and simulation of the detector response, GEANT 3 is used \cite{Brun:1994aa}.

\subsection{Event and track selection}

In order to keep the conditions of the detectors as uniform as possible, to avoid edge effects, and reject residual background collisions, the coordinate of the primary vertex along the beam axis is required to be within $\pm$10~cm from the nominal interaction point. The primary vertices are identified either using tracks reconstructed in the full central barrel or with the SPD.
The contamination from pile-up events
is reduced to a negligible level by rejecting events with multiple vertices.
Pile-up vertices identified with the SPD are required to be reconstructed using a minimum number of contributors dependent on the total number of SPD track segments (tracklets) in the event and have to be compatible with the expected collision region.
A tracklet is defined as a straight line connecting two SPD hits which points back to the primary vertex.
For the vertices identified using tracks reconstructed in the full central barrel, a minimum number of contributing tracks and a maximum $\chi^2$ per contributor for the vertex fit are required to reject fake pile-up vertices.
The events are rejected as pile-up events if they contain pile-up vertex candidates which are well separated in the $z$ direction.
The total number of events that survive the event selection is $5.4 \times 10^{8}$, corresponding to $\approx$ 85\% of all recorded collision events.

Because of the different magnetic rigidity and the 2-in-1 magnet design of the LHC, the momenta of the particle beams are different for asymmetric collision systems such as p--Pb. As a consequence, the center-of-mass system (CMS) is shifted in the laboratory frame by a rapidity offset $\Delta y = 0.465$ in the direction of the proton beam.
Primary track candidates with transverse momentum $\pt > 1.5$ GeV/$\textit{c}$, pseudorapidity $|\eta_\text{lab}| \leq 0.9$ and $-1 \leq y_\mathrm{cms} < 0$ are selected from those reconstructed both in the ITS and TPC by applying quality criteria that were optimized to ensure a good track momentum and \dEdx~resolution.

Tracks are required to have a minimum number of reconstructed space points in the TPC ($N_{\mathrm{cls}}^{\mathrm{TPC}}$) of 70 for $^3$He and 120 for $^3$H out of a maximum of 159 clusters, respectively. For $^3$H candidates, a stronger selection is used in order to reduce the contamination from other particle species.
In addition, at least two hits in the ITS ($N_{\mathrm{cls}}^{\mathrm{ITS}} \geq 2$), with at least one in the SPD, are requested. The latter requirement significantly suppresses the contribution of secondary tracks. During the data collection, the SDD was only read out for about half of the events recorded in order to maximize the data acquisition speed. To maximize the size of the data set and to unify the reconstruction of the events, the information from the SDD is not used for the current analyses, which reduces the maximum number of hits in the ITS to 4.

The quality of the track fit is quantified by the value of $\chi^{2}/N_{\mathrm{cls}}^{\mathrm{TPC}}$, which is required to be less than 4. 
In addition, the ratio of the number of reconstructed TPC clusters to the number of findable TPC clusters is required to be larger than 80$\%$. The number of findable clusters is the maximum number of geometrically possible clusters which can be assigned to a track.

The contribution from secondary tracks that are produced, e.g., by spallation in the detector material, is further suppressed by restricting the DCA to the primary vertex.
The absolute values of the DCA in the transverse plane ($\mathrm{DCA}_{\mathrm{xy}}$) and in the beam direction ($\mathrm{DCA}_{\mathrm{z}}$) are required to be smaller than 0.1 and 1 cm, respectively.

\subsection{Particle identification} \label{PID}

\begin{figure}[htb]
\centering
\includegraphics[width=0.49 \textwidth]{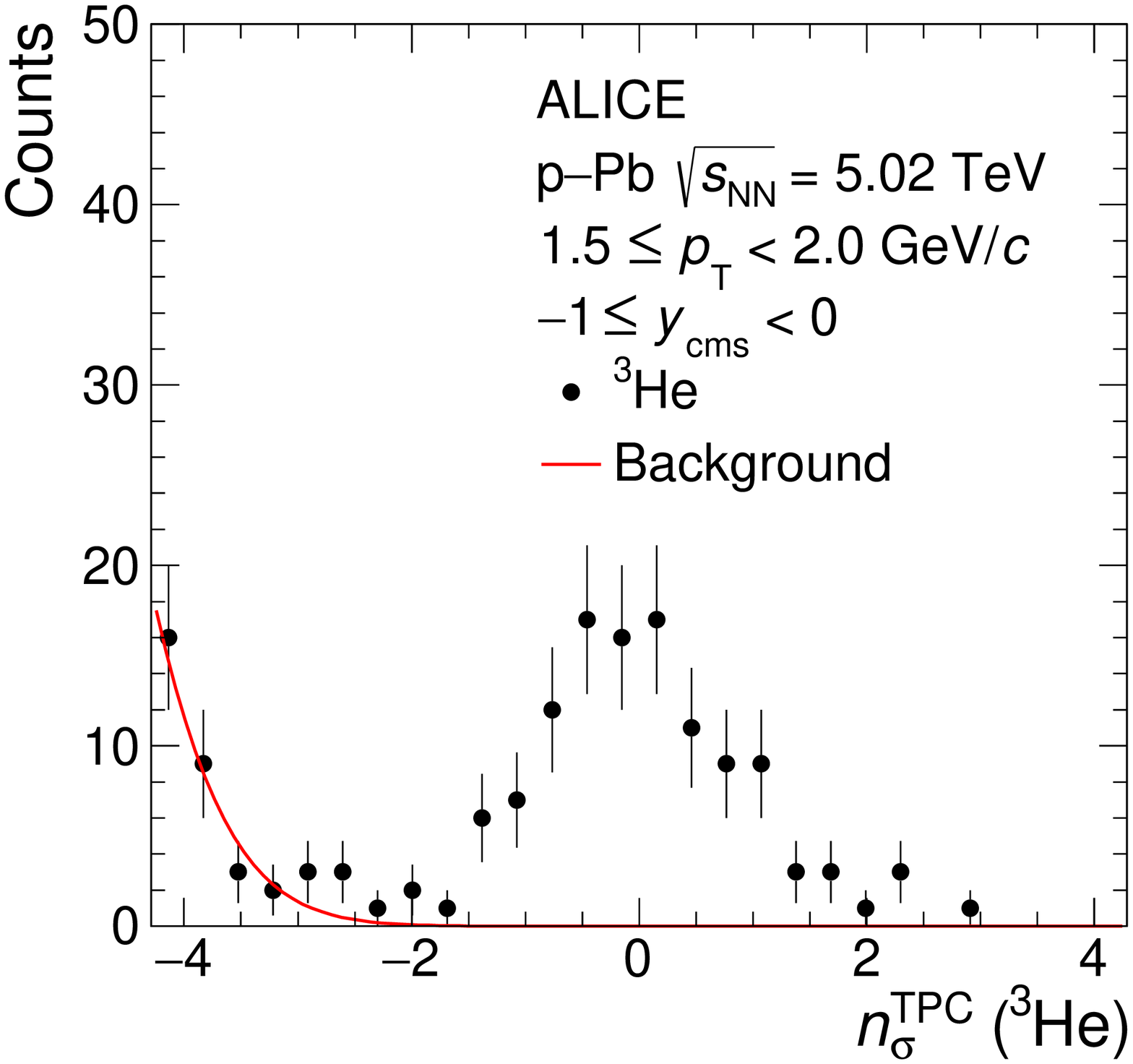}
\hfil
\includegraphics[width=0.49 \textwidth]{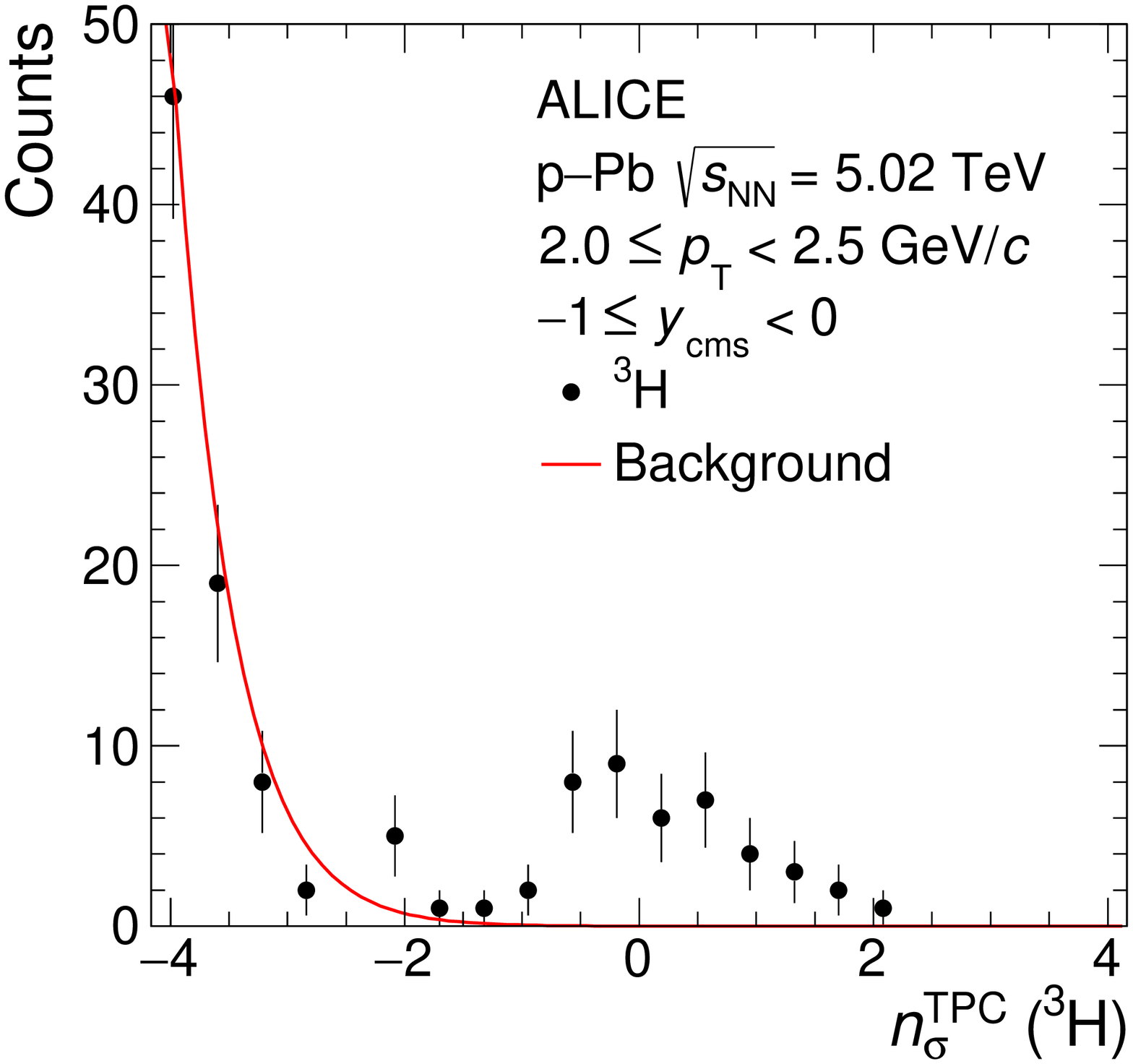}
\caption{The distribution of the specific ionization energy loss (\dEdx) in the TPC of the candidate tracks compared to the expected value for $^3$He or $^3$H ($n_\sigma^\mathrm{TPC}$) in the \pt range of 1.5 $\leq
\pt < 2.0$ GeV/$c$ and 2.0 GeV/$c$ $\le \pt < 2.5$ GeV/$c$ for $^3$He (left panel) and $^3$H (right panel), respectively. The background, which is visible as a slope on the left side of the signal, is fitted with a Gaussian function shown in red to estimate the contamination.}
\label{Figure:TPCSelection}
\end{figure}

The identification of tracks as $^{3}$He and $^{3}$H is based on the specific energy loss \dEdx measured by the TPC. For $^{3}$He, this provides excellent separation from other particle species due to the quadratic dependence of \dEdx on the particle charge. The only relevant contamination is caused by secondary $^3$H due to the similar specific energy loss in the kinetic region of $\pt < 3 $ GeV/$c$.
As shown in the left panel of \autoref{Figure:TPCSelection}, the fraction of contamination is estimated from data by fitting the slope on the left side of the $^3$He peak in the \dEdx distribution with a Gaussian function.
This contamination is found to be below 0.5\% for $^3$He, while the signal extraction of $^3\overline{\mathrm{He}}$ is not affected.
For $^{3}$H, the PID signal in the TPC contains a large background from other, more abundant particle species because $^3$H has only one elementary charge. This background is largely suppressed by applying a preselection based on the measured time of flight, which is required to be within 3$\sigma_{\mathrm{TOF}}$ from the value expected for $^3$H, where $\sigma_{\mathrm{TOF}}$ is the resolution of the time-of-flight measurement. At $\pt > 2.0$ GeV/$c$, the TOF preselection does not efficiently suppress the contamination by other particles, like electrons and pions, anymore which leads to an increasingly large contamination for higher \pt. The contamination of the signal is estimated following the same approach as for the signal extraction of $^3$He. For $^3$H ($^3\overline{\mathrm{H}}$) in the transverse momentum regions $\pt = 2-2.5$ GeV/$c$ and $\pt = 2.5-3$ GeV/$c$, the contamination is found to be $\sim 7 (9)\%$ and $\sim 34 (21)\%$, respectively.
The $^{3}$He ($^{3}$H) candidates are selected using the difference between the measured \dEdx~and the expected value for $^{3}$He ($^{3}$H), in units of the energy loss resolution of the TPC, $n_\sigma^\mathrm{TPC}$. The signal is extracted by subtracting the contamination and counting the number of candidates inside the interval $[-3\sigma,3\sigma]$.

\subsection{Secondary nuclei from material}

\begin{figure}[htb]
\centering
\includegraphics[width=0.49 \textwidth]{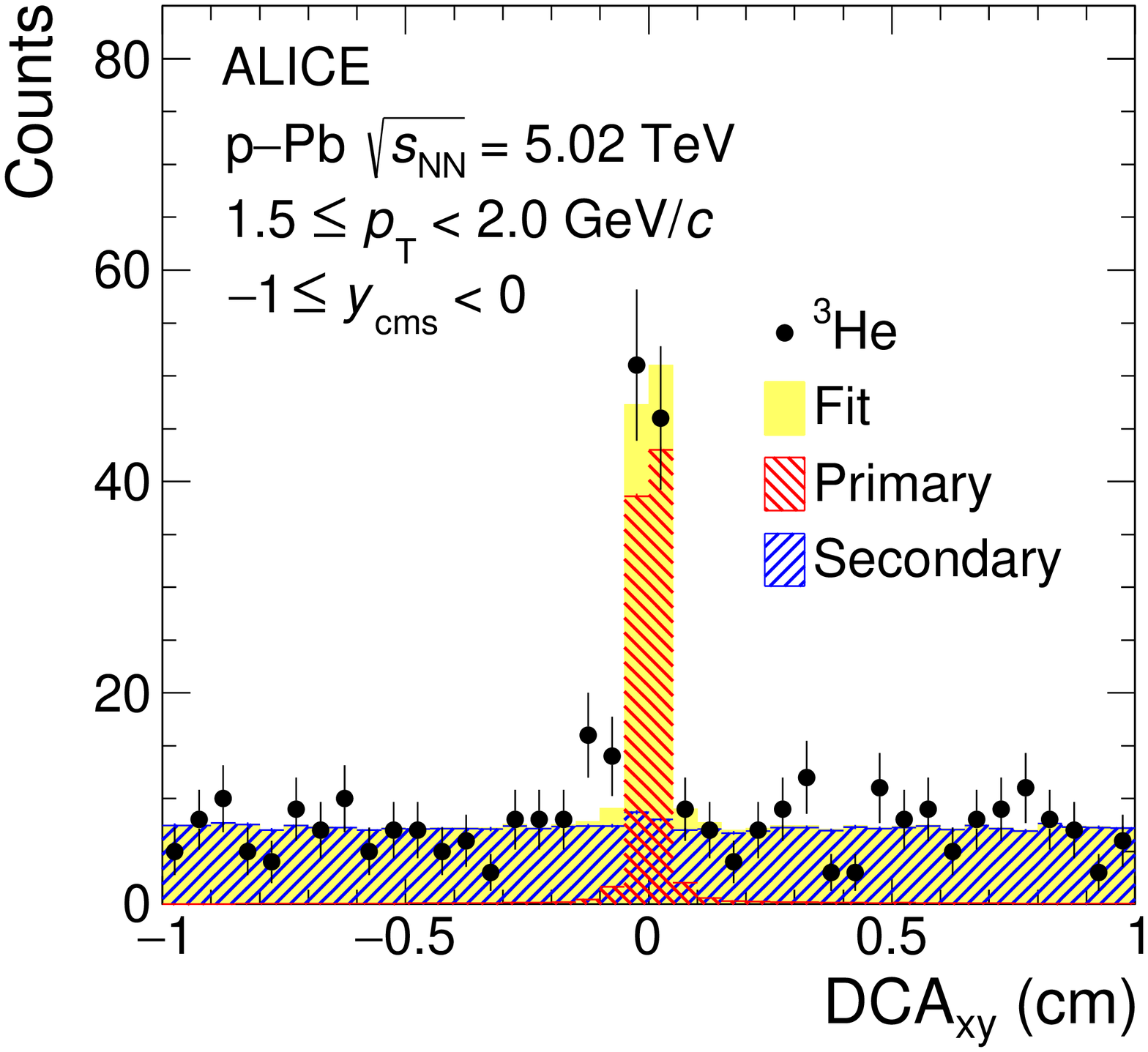}
\hfil
\includegraphics[width=0.49 \textwidth]{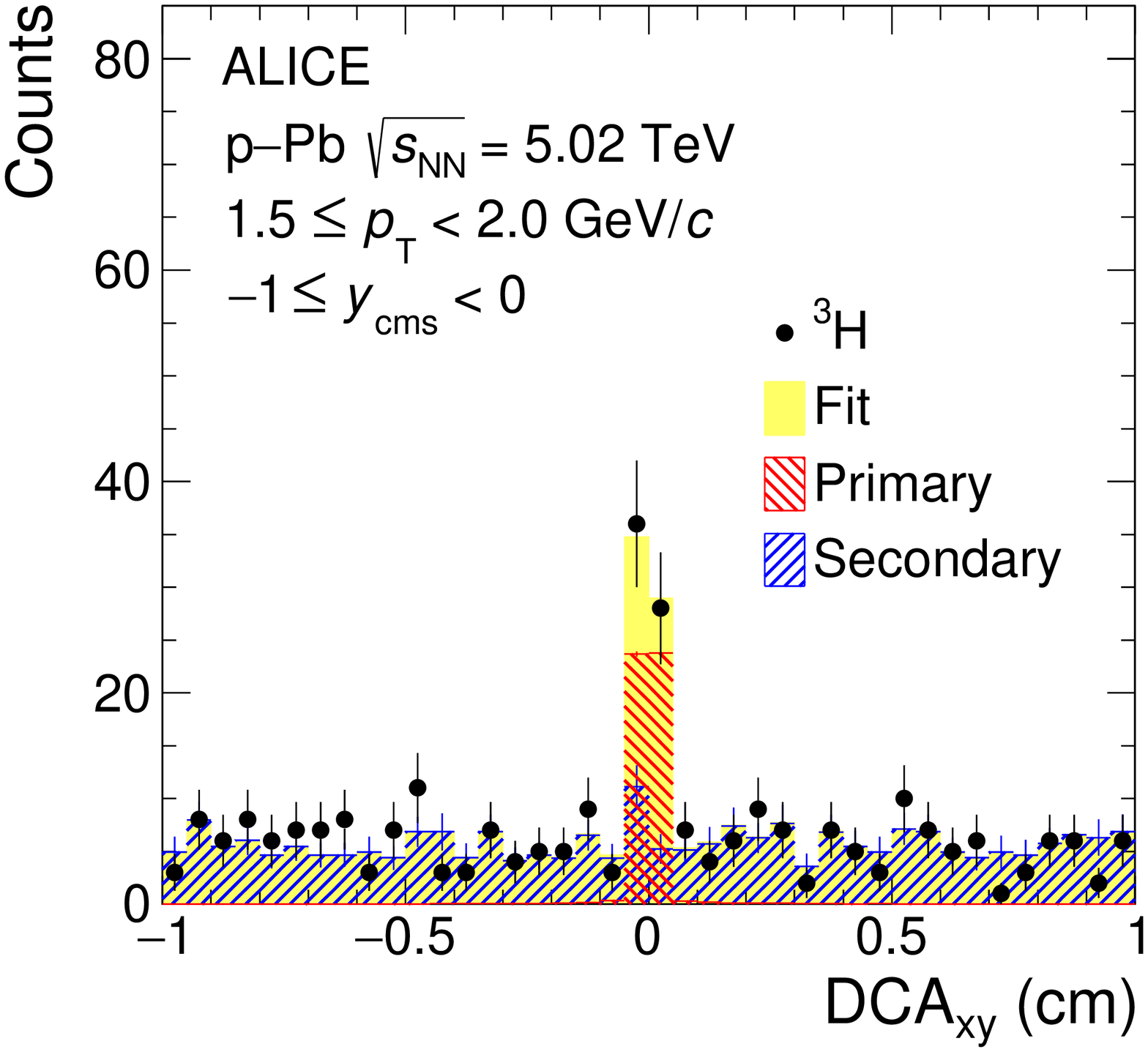}
\caption{The $\mathrm{DCA}_\mathrm{xy}$ distribution within 1.5 $\le \pt < 2.0$ GeV/$c$ is shown together with the MC template fit for $^3$He (left panel) and $^3$H (right panel). The corresponding primary and secondary contributions are also indicated.}
\label{Figure:DCA}
\end{figure}

Secondary nuclei are produced as spallation fragments in the interactions between primary particles and nuclei in the detector material or in the beam pipe. The contribution of secondary nuclei can be experimentally separated from that of primary nuclei using the $\mathrm{DCA}_\mathrm{xy}$ to the primary vertex.
The $\mathrm{DCA}_\mathrm{xy}$ distribution of primary nuclei is peaked at zero, while the one of secondary nuclei is flat over most of the $\mathrm{DCA}_\mathrm{xy}$ range and has a small peak around $\mathrm{DCA}_\mathrm{xy}=0$ cm for low \pt, as shown in \autoref{Figure:DCA}. This structure is artificially created by the tracking algorithm and is due to incorrect cluster association in the first ITS layer.

The $\mathrm{DCA}_\mathrm{xy}$ distribution of $^{3}$He in data is obtained by applying stricter PID requirements compared to those described in \autoref{PID} to ensure a pure $^{3}$He sample. In particular, the difference between the measured \dEdx and the expected average for $^{3}$He is required to be in the range $[-2\sigma, 3\sigma]$ for $p_\text{T} < 2$ GeV/$\textit{c}$ and in the range $[-2.5\sigma, 3\sigma]$ for $2 < p_\text{T} < 2.5$ GeV/$\textit{c}$.
The remaining contamination is at maximum 0.1\% for $^3$He and 1.2\% for $^3$H for $\pt < 2.5$ GeV/$c$.

The fraction of primary nuclei is obtained by a two-component fit to the measured $\mathrm{DCA}_\mathrm{xy}$ distribution, one for the signal and the other for the secondaries. The distribution of both components is obtained from Monte Carlo simulations.
Because of the lack of secondary $^3$He in the MC simulation, the distributions of secondary deuterons are used as a proxy. For a given \pt, the template of deuterons at $p_\text{T}/2$ is used to compensate for the charge difference. The different multiple scattering for deuterons and $^3$He has a negligible impact on the $\mathrm{DCA}_\mathrm{xy}$ distribution. This is confirmed by comparing the $\mathrm{DCA}_\mathrm{xy}$ distributions of antideuteron and $^3\overline{\mathrm{He}}$ candidates in data for the same interval of transverse rigidity ($\pt/q$).
For $p_\text{T} > 2.5$~GeV/$\textit{c}$, the $\mathrm{DCA}_\mathrm{xy}$ distributions of $^{3}$He and $^{3}$H are well reproduced using only the template for primary nuclei, which implies that the fractions of secondary $^{3}$He and $^{3}$H are negligible or below the sensitivity of this measurement.
The fraction of primary nuclei is calculated in the range $|\mathrm{DCA}_\mathrm{xy}| \leq 0.1$~cm. The resulting values are summarized in \autoref{table:PrimaryFraction}. 

\begin{table}[htb]
\centering
\caption{The primary fraction calculated for $^3$He and $^3$H with its uncertainty.}
\label{table:PrimaryFraction}
\begin{tabular}{lcc}
\hline 
\pt (GeV/$c$) & $^3$He & $^3$H \\ 
\hline
$1.5 - 2.0$ & $(73 \pm 1)\%$ & $(65 \pm 1)\%$ \\
$2.0 - 2.5$ & $(94.5 \pm 0.2)\%$ & $(97 \pm 1)\%$ \\
above 2.5 & $ 100\%$ & $ 100\%$ \\
\hline
\end{tabular}
\end{table} 

The fractions of primary nuclei calculated in different multiplicity intervals are consistent with those calculated for the minimum-bias data sample within uncertainties. Because of the limited number of $^3$He candidates, the fit is highly unstable for the lowest multiplicity.
Therefore, the primary fraction is calculated using the minimum-bias data sample and used to correct the spectra in all the multiplicity intervals.

\subsection{Efficiency and acceptance}

\begin{figure}[htb]
\centering
\includegraphics[width=0.7 \textwidth]{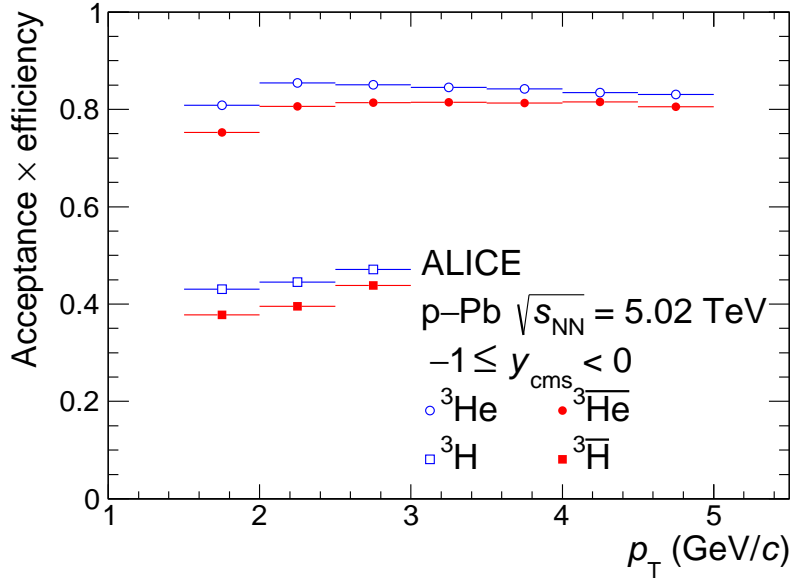}
\caption{The acceptance $\times$ efficiency as a function of \pt is shown for $^3\overline{\mathrm{He}}$ and  $^3\text{He}$ as well as for $^3\overline{\mathrm{H}}$ and $^3$H.}
\label{Figure:Efficiencies}
\end{figure}

The product of the acceptance and the efficiency is calculated as the ratio between reconstructed and generated primary nuclei in the MC simulation within $-1\le y_\mathrm{cms} < 0$ and 1 $\le \pt < 5$ GeV/$c$ or 1 $\le \pt < 3$ GeV/$c$ for (anti-)$^{3}$He and \text{(anti-)}$^{3}$H, respectively. 
The same track selection criteria that are used in data are applied to the reconstructed particles in the simulation. The acceptance $\times$ efficiency of (anti-)$^{3}$H and (anti-)$^{3}$He are shown in \autoref{Figure:Efficiencies} as a function of \pt. 

The efficiency for (anti-)$^3$H is lower compared to that of (anti-)$^{3}$He due to the larger number of TPC clusters required and the additional requirement of a hit in the TOF detector. The latter implies the crossing of the additional material between the TPC and the TOF detector. Nuclear absorption and multiple Coulomb scattering reduce the TPC-TOF matching efficiency, leading to a lower efficiency for $^3$H. Furthermore, the efficiency and the acceptance of the TOF detector have to be taken into account. The efficiency for the antinuclei is reduced compared to the one for the nuclei due to annihilation processes with the beam pipe and the detector material.

\subsection{Feed-down from hypertriton} \label{Feed-down}

The transverse momentum distribution of (anti-)$^3$He and (anti-)$^3$H contains a contribution from weak decays of (anti-)hypertriton, $^3_\Lambda \mathrm{H} \rightarrow\,^3\text{He} + \pi^-$ and $^3_\Lambda \mathrm{H} \rightarrow\,^3\mathrm{H} + \pi^0$ and charge conjugates. The \discretionary{(anti\hbox{-})}{hypertriton}{(anti\hbox{-})hypertriton} represents the only relevant source of feed-down at LHC energies. The goal of this study is the measurement of primary (anti-)$^{3}$He and (anti-)$^{3}$H produced in the collision. For this reason, the contribution of secondary (anti-)$^{3}$He and (anti-)$^{3}$H produced in weak decays of (anti-)hypertriton, estimated using the simulations, is subtracted from the inclusive \pt distribution. 

The fraction of secondary (anti-)$^3$He from \discretionary{(anti\hbox{-})}{hypertriton}{(anti\hbox{-})hypertriton} decays is given by:

\begin{align}
f_\text{feed-down}(\pt) 
 & = \tfrac{\epsilon_\text{feed-down}(\pt)}{\epsilon_{^3\text{He}}(\pt)} \; \text{BR} \; \tfrac{^3_\Lambda \text{H}}{^3\text{He}} 
\label{Eq:Feed-down}
\end{align}

where $\epsilon_\text{feed-down}$ and $\epsilon_{^3\text{He}}$ are the reconstruction efficiencies of secondary $^{3}$He from \discretionary{(anti\hbox{-})}{hypertriton}{(anti\hbox{-})hypertriton} decays and primary $^{3}$He, respectively.
The DCA selection introduced to suppress the secondaries from the interaction with material also reduces the reconstruction efficiency for feed-down $^3$He by about 40\% compared to the one for primary $^3$He.
    BR denotes the branching ratio of the decay of $^3_\Lambda$H into $^3$He which is about 25\% \cite{Kamada:1997rv}. The (anti-)$^3_{\Lambda}$H-to-(anti-)$^{3}$He ratio is extrapolated to the analyzed multiplicity class from those measured as a function of d$N$/d$\eta_\mathrm{lab}$ in Pb--Pb collisions at \energy~= 2.76 TeV \cite{Adam:2015yta}.

An upper limit for this contribution to $^3$H is evaluated as half of the contribution for $^3$He since the branching ratio of the two-body decay with neutral daughters is half the one with charged particles \cite{Kamada:1997rv}.
The measured \pt spectra of (anti-)$^3$He and (anti-)$^3$H are corrected for the fraction of secondary (anti-)$^3$He and (anti-)$^3$H from (anti-)hypertriton decays, which is estimated to be about $3.7\%$ and $1.9\%$, respectively.

\section{Systematic uncertainties}

The main sources of systematic uncertainties on the (anti-)$^{3}$He and (anti-)$^{3}$H yields are summarized in \autoref{Tabel:SystematicUnertainties} and discussed in the following.
The procedures used for the evaluation of the systematic uncertainties quantify effects due to residual discrepancies between the data and the MC used to evaluate the reconstruction efficiency.
The total systematic uncertainties are calculated as the sum in quadrature of the individual contributions assuming that they are uncorrelated.

\begin{table}[b]
\centering
\caption{Summary of the individual contributions to the total systematic uncertainty in the lowest and highest \pt interval measured for $^3$He and $^3$H. The values for the antinuclei are shown in the parentheses if they differ from the corresponding values for the nuclei.}
\label{Tabel:SystematicUnertainties}
\begin{tabular}{lcccc}
\hline
Particle & \multicolumn{2}{c}{$^3\text{He}$ ($^3\overline{\mathrm{He}}$)} & \multicolumn{2}{c}{$^3\mathrm{H}$ ($^3\overline{\mathrm{H}}$)} \\
\hline
\pt interval (GeV/$c$) & 1.5--2.0 & 4.5--5.0 & 1.5--2.0 & 2.5--3.0 \\
\hline
Tracking & 4\% & 5\% & 5\% & 5\% \\
PID \& contamination & 3\% (1\%) & 1\% & 3\% (5\%) & 20\% (30\%) \\
Primary fraction estimation & 9\% (negl.) & negl. & 6\% (negl.)  & 3\% (negl.)  \\
Material budget & 0.3\% (0.5\%) & 0.2\% (0.5\%) & 2.0\% (3.4\%) & 0.7\% (1.3\%) \\ 
Hadronic cross section & 9\% (6\%) & 1\% (2\%) & 2\% (8\%) &  negl. (11\%)\\
Feed-down & 1.5\% & 1.5\% & 0.8\% & 0.8\% \\
\hline 
Total systematic uncertainty & 13\% (7\%) & 5\% (6\%) & 9\% (12\%) & 20\% (32\%)\\
\hline
\end{tabular}
\end{table}  

The systematic uncertainty related to track reconstruction contains contributions coming from the different matching efficiencies between ITS and TPC for $^3$He and $^3$H, and between TPC and TOF for $^3$H in data and MC and a contribution due to the track selection criteria used in the analysis.
The latter is estimated by varying the track selection criteria, both for data and in the MC for the efficiency calculation. For each transverse momentum interval, the systematic uncertainty is given by the root mean square (rms) of the spread of data points, each corresponding to a given track selection. The corresponding uncertainty is found to be $4-5\%$.
The uncertainties due to the different ITS-TPC and TPC-TOF matching efficiencies are both about 1$\%$. 
The total tracking systematic uncertainty is obtained as the sum in quadrature of each contribution and is found to be about $4-5\%$ for both $^3$He and $^3$H, independent of \pt.

The uncertainty from the particle identification is estimated by varying the fit function used to describe the contamination and by changing the fitting ranges in the TPC and TOF for the signal extraction as well as for the evaluation of the contamination. The latter has only a minor effect on the uncertainty for $^3$He due to its clear separation from other charged particles. In contrast, the effect of the contamination on $^3$H is much larger because the separation from other charged particles, which are much more abundant, decreases with increasing \pt. An exponential function is also used, besides a simple Gaussian, to describe the $^3$H contamination in the $^3$He signal and the contamination in the $^3$H signal. The resulting difference is included in the systematic uncertainty due to the PID and the contamination which amounts to maximally 3\% and 30\% for $^3$He and $^3\overline{\mathrm{H}}$, respectively.

The systematic uncertainty associated with the fraction of primary nuclei contains three sources: the uncertainty of the template fit, the stability against including more secondaries, and the possible bias of the templates used.
For the latter contribution, a Gaussian function is used to describe the DCA$_{\mathrm{xy}}$ distribution of secondary nuclei, while the distribution of antinuclei is used as a template for the primary nuclei.
The parameters of the Gaussian function are obtained by fitting the DCA$_{\mathrm{xy}}$ distribution excluding the region $|\mathrm{DCA}_{\mathrm{xy}}| \leq 0.1$ cm. 
The fraction of primary nuclei is calculated using two methods: In one case, the template for primary nuclei and the Gaussian background are used, and in the other case, only the Gaussian function is used. In addition, the primary fraction is calculated using MC templates from secondary $^3$H scaled in the same way as the deuteron templates.
The maximum difference between the fraction of primary nuclei obtained from these methods is divided by $\sqrt{12}$ to estimate the systematic uncertainty.
The stability of the primary fraction correction is tested by varying the DCA selection and, thus, varying the number of secondary nuclei taken into account. The primary fraction should adjust accordingly. This uncertainty is evaluated using an rms approach.
The total uncertainty linked to the primary fraction estimate is given by the sum in quadrature of the three components. It is found to be at maximum 9\% for $^3$He and 6\% for $^3$H and follows a decreasing trend with \pt.

The material budget of the detector, i.e. the thickness up to the middle of the TPC, expressed in units of the radiation length, is known with a relative uncertainty of 4.5\% \cite{Abelev:2014ffa}, which leads to an uncertainty on the reconstruction efficiency. The impact of this uncertainty on the results is studied by evaluating the relative uncertainty on the reconstruction efficiency using a dedicated MC production with 4.5\% higher or lower material budget. The relative uncertainty $\sigma_\text{material budget}$ is calculated via

\begin{align}
\sigma_\text{material budget}(\pt) = \frac{\epsilon_\text{max}(\pt) - \epsilon_\text{min}(\pt)}{2\;\epsilon_\text{default}(\pt)},
\end{align}

where $\epsilon_\text{max}$ and $\epsilon_\text{min}$ are the largest and the smallest efficiencies obtained in a given \pt interval. $\epsilon_\text{default}$ denotes the efficiency calculated with the default material budget. The effect is larger for $^3$H than for $^3$He because of the additional detector material which has to be taken into account when including the TOF detector in the analysis.

To evaluate the reconstruction efficiency GEANT3 was used to propagate the particles through the detectors. In the GEANT3 version used for this analysis, an empirical parametrization of the antideuteron absorption cross section, based on the measurements carried out at the U-70 Serpukhov accelerator \cite{Denisov:1971im, Binon:1970yu}, is used. Elastic scattering processes are not taken into account by this description. In GEANT4 \cite{Agostinelli:2002hh}, a Glauber model based on the well-measured total and elastic $\mathrm{p}\overline{\mathrm{p}}$ cross section is implemented \cite{Uzhinsky:2011zz}.
Thus, the systematic effect due to the incomplete knowledge about the hadronic interaction cross section of nuclei is evaluated using half of the relative difference between the reconstruction efficiency evaluated with GEANT3 and GEANT4. This contribution is found to be smaller than 12\% for (anti-)$^3$He and (anti-)$^3$H.

The last contribution to the systematic uncertainties is the feed-down from weak decays of hypertritons. In \autoref{Feed-down}, this contribution is estimated using an extrapolation of the measured $^3_\Lambda$H-to-$^3$He ratio assuming a linear trend with the charged particle multiplicity. This extrapolation is repeated after shifting the measured data points up and down by their uncertainties such that the resulting slope is maximal or minimal. The resulting maximal or minimal $^3_\Lambda$H-to-$^3$He ratios are used to calculate the relative uncertainty on the feed-down contribution given by the difference of the maximum (6.3\%) and the minimum (1.1\%) feed-down contribution divided by $\sqrt{12}$. The corresponding contribution to the total systematic uncertainty is found to be 1.5\% for $^3$He and 0.75\% $^3$H.
\section{Results}

\subsection{Transverse momentum spectra}

The production yields of (anti-)$^3$He and (anti-)$^3$H as a function of \pt are obtained by multiplying the observed number of candidate nuclei after the statistical subtraction of the contamination ($N_\text{obs}$) with the fraction of primary nuclei ($f_\text{prim}$) and correcting for the reconstruction efficiency ($\epsilon$) in each \pt interval. Afterward, the feed-down nuclei from hypertriton decays are subtracted. The corrected number of observed nuclei is divided by the number of selected events ($N_\text{events}$), the width of the transverse momentum bins ($\Delta \pt)$ and the rapidity interval ($\Delta y$).
The resulting \pt-differential yields of (anti-)$^3$He and (anti-)$^3$H correspond to the ones in $\mathrm{INEL}>0$ events because the signal and event loss due to the event selection and the trigger were found to match within less than 1\% and thus no corrections are applied.
The event class $\mathrm{INEL}>0$ contains events in which the colliding ions interact via inelastic collisions and at least one charged particle could be measured in $|\eta| < 1$.

\begin{align}
\frac{{\rm d}^2N}{{\rm d}y \, {\rm d}\pt} = \frac{1}{\Delta y \, \Delta \pt \, N_\text{events}} \frac{f_\text{prim}(\pt) \, N_\text{obs}(\pt) }{\epsilon(\pt)} (1 - f_\text{feed-down}(\pt))
\end{align}

\begin{figure}[htb]
\centering
\includegraphics[width=0.8\textwidth]{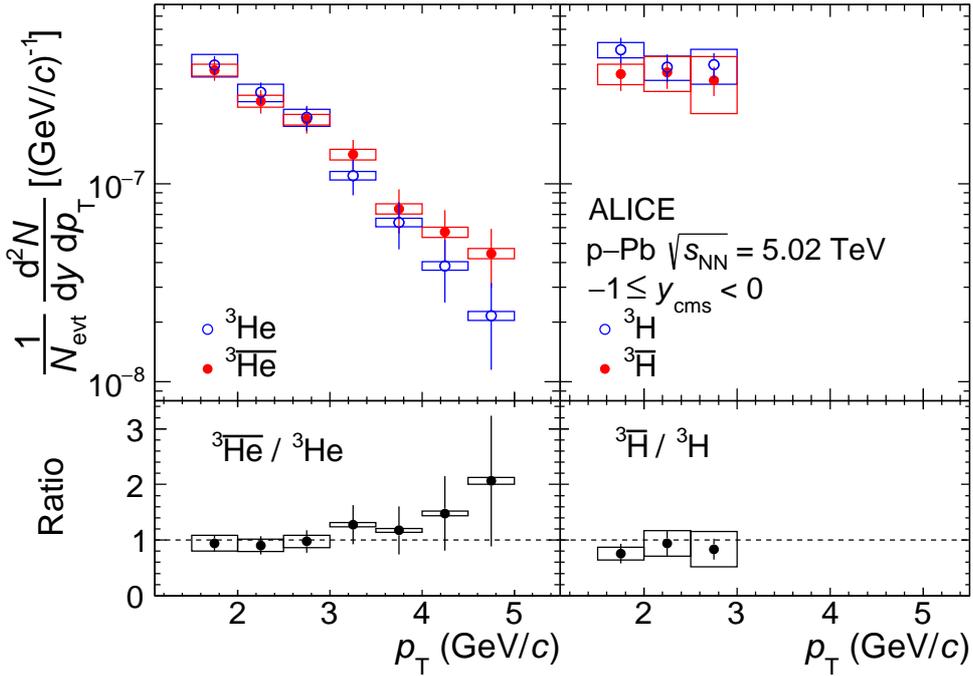}
\caption{\pt spectra of (anti-)$^3$He (left) and (anti-)$^3$H (right) measured in $\mathrm{INEL}>0$ p--Pb collisions at $\energy = 5.02$ TeV. The bottom panels show the corresponding antiparticle-to-particle ratios as a function of \pt. Statistical and systematic uncertainties are indicated by vertical bars and boxes, respectively.}
\label{Figure:Yield_Mult_Int}
\end{figure}

The minimum-bias \pt-differential yields of (anti-)$^3$He and (anti-)$^3$H measured in p--Pb collisions at $\energy = 5.02$ TeV and the corresponding antiparticle-to-particle ratios are shown in \autoref{Figure:Yield_Mult_Int}.
The antiparticle-to-particle ratio is consistent with unity within uncertainties. This indicates that matter and antimatter are produced in equal amounts in p--Pb collisions at $\energy = 5.02$ TeV. This is also observed for other light (anti)nuclei in different collision systems and center-of-mass energies at the LHC \cite{Acharya:2017fvb, Adam:2015vda}.
For the calculation of the systematic uncertainty of the antiparticle-to-particle ratio, the systematic uncertainties on the spectra were propagated, taking into account that some of them are correlated between antiparticles and particles, i.e. the uncertainty linked to the tracking, the material budget, and the feed-down.

\begin{figure}[htb]
\centering
\includegraphics[width=0.7 \textwidth]{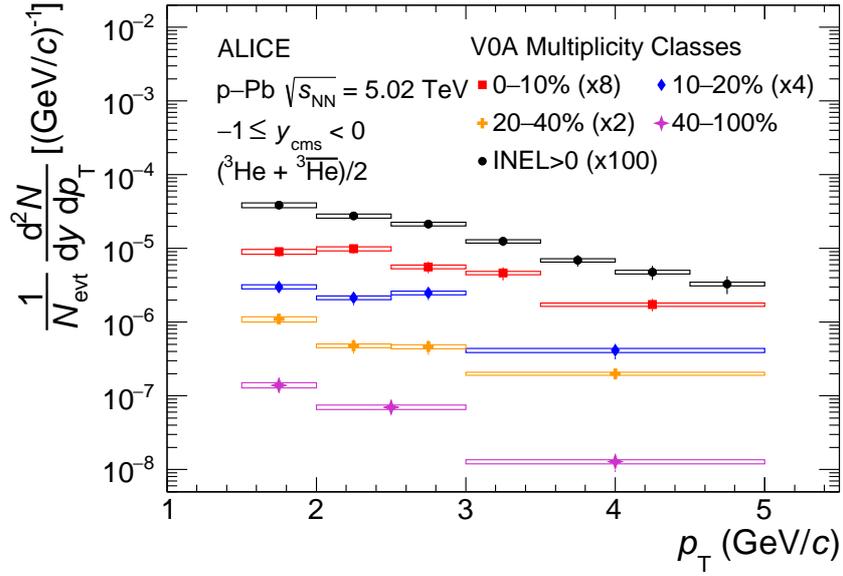}
\caption{Transverse momentum spectra obtained from the average of $^3$He and $^3\overline{\mathrm{He}}$ for four different multiplicity classes and $\mathrm{INEL}>0$ events in p--Pb collisions at $\energy = 5.02$ TeV. Different scaling factors are used for better visibility. Statistical and systematic uncertainties are indicated by vertical bars and boxes, respectively.}
\label{Figure:Summary_Yields}
\end{figure}

The \pt spectra, which are the average of $^3$He and $^3\overline{\mathrm{He}}$, are summarized in \autoref{Figure:Summary_Yields} for different multiplicity classes and $\mathrm{INEL}>0$ events.
The \pt spectra of $^3$He and $^3\overline{\mathrm{He}}$, as well as of $^3$H and $^3\overline{\mathrm{H}}$ have to be extrapolated to the unmeasured regions in order to obtain the integrated yield (\dNdy).
For the extrapolation, the measured \pt spectra are fitted with the following functional forms: \pt-exponential, $m_\mathrm{T}$-exponential, Boltzmann, Bose--Einstein, and Fermi--Dirac function.

The extrapolated yield is calculated by integrating each of these functions outside the measured \pt range and taking the average. The result is added to the integral of the measured spectrum to obtain the total \pt-integrated yield. 
For the calculation of the statistical uncertainty on the yield, the transverse momentum spectrum is modified by shifting the data points for different transverse momentum bins independently by random numbers with Gaussian distributions centered around the measured values with a width given by the statistical uncertainties. In addition, the extrapolated yields at \pt below and above the measured range are varied following a Gaussian function centered at the default value with a width given by the uncertainty on the extrapolated yield.
The standard deviation of the distribution of measured yields determines the statistical uncertainty for each functional form fitted.

For the systematic uncertainty of the total yield, the uncertainties which are correlated in \pt, i.e. the material budget, the hadronic cross section, feed-down uncertainty, and the uncertainty linked to the estimation of the primary fraction, are treated separately from the remaining uncertainties for each of the functional forms. The resulting contribution is evaluated as the average difference between the default value and the yield obtained by shifting the measured points up or down by the correlated part of the systematic uncertainties.
The remaining part of the total uncertainty, i.e. the uncertainty linked to the track selection, PID, and contamination, is partially uncorrelated between \pt bins. Therefore, the Gaussian sampling procedure is also used to evaluate the contributions of these sources to the systematic uncertainty of the \pt-integrated yield.
The contribution for each functional form is given by the sum in quadrature of the uncorrelated and the correlated uncertainty.
To obtain the total systematic uncertainty on the integrated yield, the average of the contributions from the different functional forms is calculated and added in quadrature to the uncertainty given by the spread of the values obtained with the different functional forms.
The latter is calculated as the difference of the maximum and the minimum yield divided by $\sqrt{12}$. The extrapolated fraction of the integrated yield below and above the measured \pt interval is summarized in \autoref{Table:Extrapolation}.

\begin{table}[h]
\begin{center}
\centering
\caption{Fraction of extrapolated yields below and above the measured \pt interval.}
\label{Table:Extrapolation}
\begin{tabular}{llcc}
$^3$He & Event class & $\pt < 1.5$ GeV$/c$  & $\pt > 5$ GeV$/c$ \\
\hline
& 0--10\% & $(39 \pm 5)\%$ & $(2.4 \pm 0.8)\%$\\
& 10--20\% & $(46 \pm 7)\%$ & $(0.8 \pm 0.4)\%$\\
& 20--40\% & $(38 \pm 7)\%$ & $(2 \pm 1)\%$\\
& 40--100\% & $(55 \pm 8)\%$ & $(0.3 \pm 0.2)\%$\\
& $\mathrm{INEL}>0$ & $(43 \pm 5)\%$ & $(1.4 \pm 0.4)\%$\\
\hline
\hline
$^3$H & Event class & $\pt < 1.5$ GeV$/c$  & $\pt > 3$ GeV$/c$ \\
\hline
& $\mathrm{INEL}>0$ & $(24 \pm 13)\%$ & $(38 \pm 16)\%$\\
\hline
\end{tabular}
\end{center}
\end{table}

\begin{figure}[h]
\centering
\includegraphics[width=0.49 \textwidth]{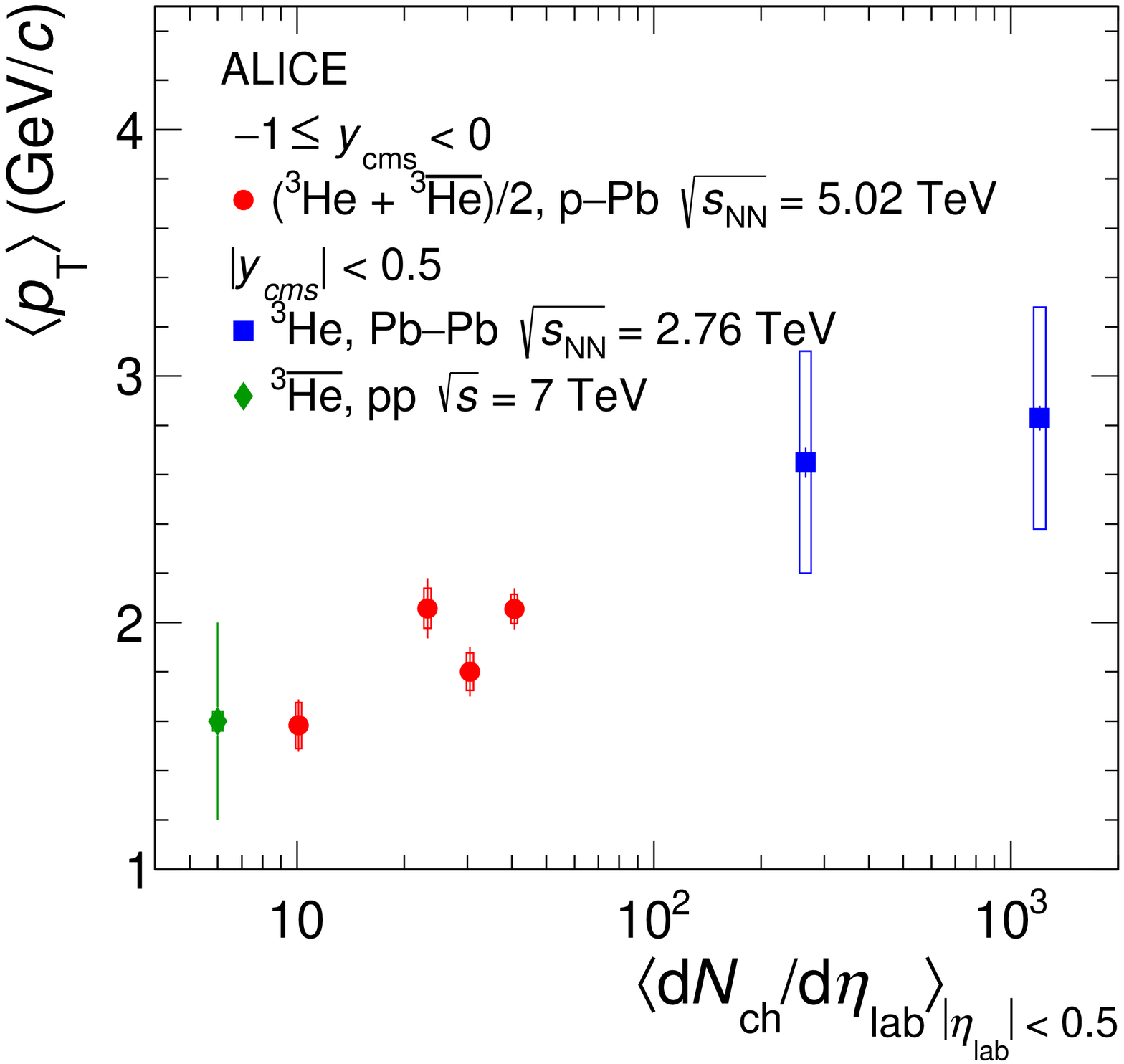}
\includegraphics[width=0.49 \textwidth]{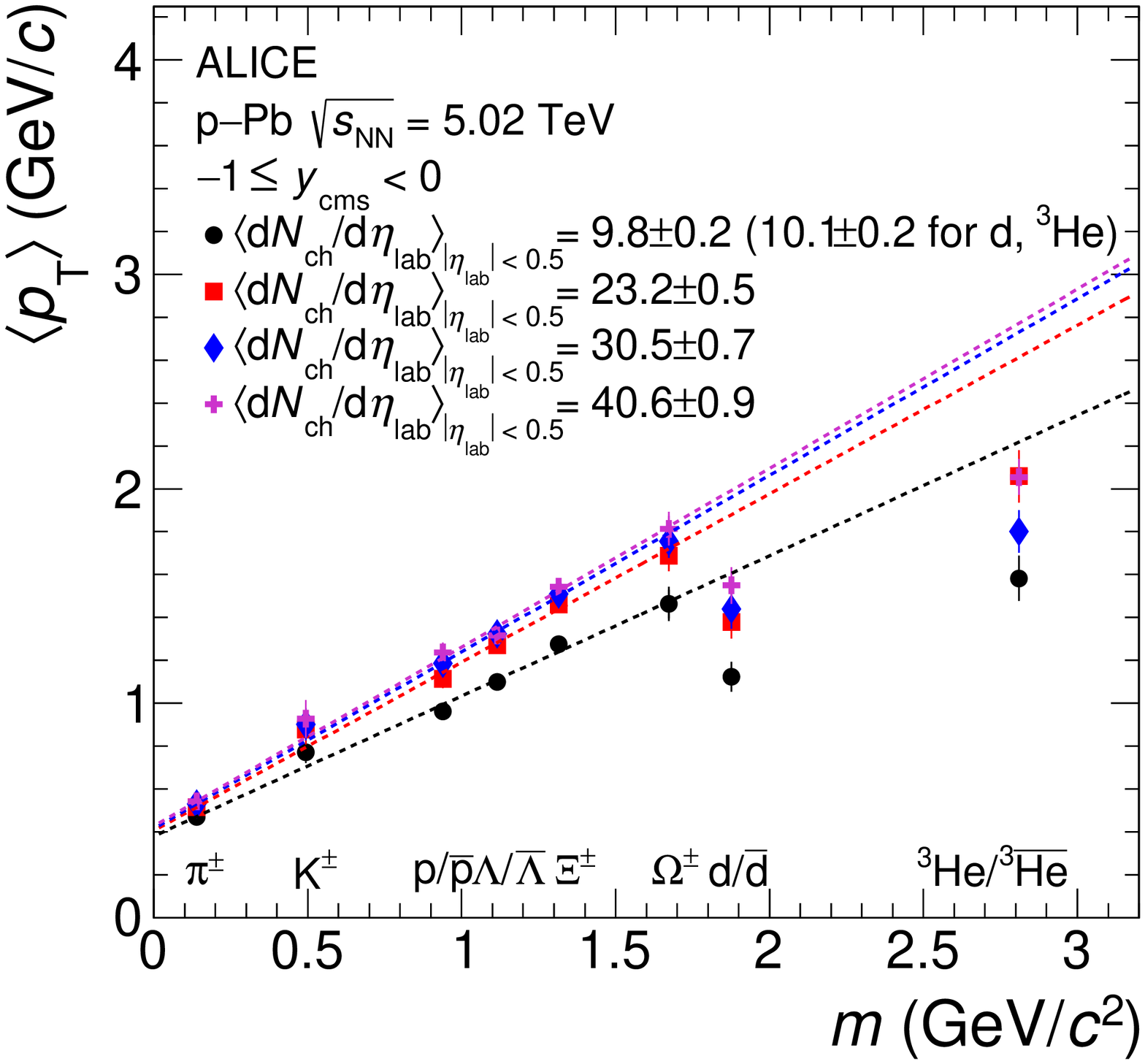}
\caption{Left: Mean transverse momentum of (anti-)$^3$He as a function of the mean charged-particle multiplicity density in p--Pb collisions at $\energy = 5.02$ TeV. Statistical and systematic uncertainties are indicated by vertical bars and boxes, respectively. The published results from pp \cite{Acharya:2017fvb} and Pb--Pb \cite{Adam:2015vda} collisions are shown with diamonds and rectangles, respectively.
Right: Mean transverse momentum measured in p--Pb collisions at $\energy = 5.02$ TeV as a function of the particle mass is shown for different mean charged-particle multiplicity densities. The linear scaling with the mass found for the results for $\pi$, K, p, $\Lambda$ \cite{Abelev:2013haa}, $\Xi$, and $\Omega$ \cite{Adam:2015vsf} is indicated by dashed lines. The deuteron $\langle p_{\mathrm{T}}\rangle$ is taken from \cite{Acharya:2019rys}.}
\label{Figure:MeanPt}
\end{figure}

Based on the extrapolation, the mean transverse momenta ($\langle p_{\mathrm{T}}\rangle$) of the average $^{3}$He and $^{3}\overline{\mathrm{He}}$ yields are calculated for the different multiplicity classes. The statistical and systematic uncertainties on $\langle p_{\mathrm{T}}\rangle$ are calculated in a similar way as for the integrated yield.
The result is shown and compared with the $\langle p_{\mathrm{T}}\rangle$ measured in pp collisions at $\sqrt{s} = 7$ TeV \cite{Acharya:2017fvb} and in Pb--Pb collisions at $\energy = 2.76$ \cite{Adam:2015vda} in the left panel of \autoref{Figure:MeanPt}.
The $\langle p_{\mathrm{T}}\rangle$ measured in p--Pb collisions increases with the mean charged-particle multiplicity density, connecting the measured results in pp \cite{Acharya:2017fvb} and Pb--Pb collisions \cite{Adam:2015vda} in a smooth way.
This indicates a hardening of the \pt spectra with increasing mean charged-particle multiplicity density, which might be caused by production in jets \cite{Acharya:2019mzb} or by collective expansion effects \cite{Abelev:2013haa}. The latter would also result in a shift of the maximum of the \pt distribution, which cannot be observed in the present measurements due to the limited statistical precision.

If the system evolves following a hydrodynamic expansion, the mean transverse momenta of different particle species should follow a mass ordering, as a result of the radial flow.
In the right panel of \autoref{Figure:MeanPt}, the $\langle p_{\mathrm{T}}\rangle$ as a function of the particle mass is shown for different mean charged-particle multiplicity densities.
For similar \dndeta, a clear mass ordering is observed for the different particle species. 
The measurements for the nuclei prefer a scaling which does not follow the same linear trend as the results for $\pi$, K, p, $\Lambda$ \cite{Abelev:2013haa}, $\Xi$, and $\Omega$ \cite{Adam:2015vsf}.

\subsection{Ratio to protons}

\begin{figure}[htb]
\centering
\includegraphics[width=0.7\textwidth]{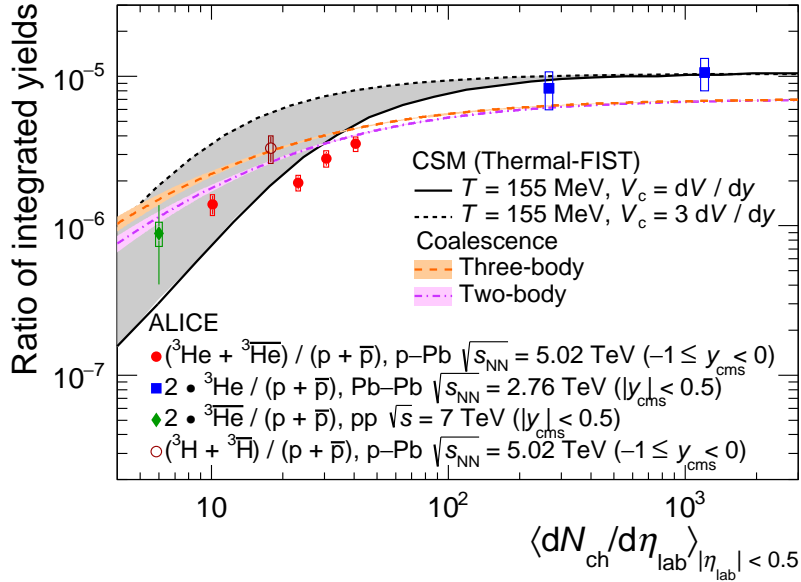}
\caption{$^3$He/p ratio in pp, p--Pb, and Pb--Pb collisions \cite{Acharya:2017fvb,Adam:2015vda} as a function of the mean charged-particle multiplicity density, together with the $^3$H/p ratio. Statistical and systematic uncertainties are indicated by vertical bars and boxes, respectively. The expectations for the canonical statistical hadronization model (Thermal-FIST \cite{Vovchenko:2018fiy}) and two coalescence approaches \cite{Sun:2018mqq} are shown. For the thermal model, two different values of the correlation volume are displayed.
The uncertainties of the coalescence calculations, which are due to the theoretical uncertainties on the emission source radius, are denoted as shaded bands.}
\label{RatioToProton}
\end{figure}

The ratio of the integrated yields of (anti-)$^{3}$He to those of (anti)protons ($^3$He/p) is calculated for the four multiplicity classes used in this analysis, while the yield ratio of (anti-)$^{3}$H to (anti)protons ($^3$H/p) is calculated for $\mathrm{INEL}>0$ events.
The \pt-integrated proton yields are taken from Ref.~\cite{Abelev:2013haa}. The $^{3}$He/p and the $^{3}$H/p ratios are shown as a function of the mean charged-particle multiplicity density in \autoref{RatioToProton}, together with the ones from pp collisions at $\sqrt{s} = 7 $ TeV \cite{Acharya:2017fvb} and from Pb--Pb collisions at $\energy =$ 2.76 TeV \cite{Adam:2015vda}.
The measured ratio is larger in Pb--Pb collisions with respect to pp collisions.
The value measured in central Pb--Pb collisions is consistent with the prediction of the grand canonical version of the SHM \cite{Andronic:2010qu,Andronic:2017pug}.
The results obtained in p--Pb collisions show an increasing trend as a function of the mean charged-particle multiplicity density and indicate a smooth transition from pp to Pb--Pb collisions.

In \autoref{RatioToProton}, the data are compared to the expectations from the Canonical Statistical hadronization Model (CSM)\cite{Vovchenko:2018fiy} and two coalescence approaches \cite{Sun:2018mqq}.
The trend observed in the data can be qualitatively reproduced over the full multiplicity range using the CSM approach, which is based on exact conservation of charges across the correlation volume $V_{\mathrm{c}}$ \cite{Vovchenko:2018fiy}.
The predictions were calculated using a temperature $T=155$ MeV and a correlation volume extending across one unit ($V_\mathrm{c} = \text{d}V/\text{d}y$) and three units ($V_\mathrm{c} = 3\,\text{d}V/\text{d}y$) of rapidity. The temperature value is constrained by the ratio measured in Pb--Pb collisions \cite{Adam:2015vda}. It is very close to the chemical freeze-out temperature which results in the best description by the grand canonical SHM \cite{Andronic:2017pug} of the ALICE measurements of the integrated yields of particles measured in most-central Pb--Pb collisions. For the mean charged-particle multiplicity density region covered by the results obtained in  Pb--Pb collisions, the CSM has reached the grand canonical limit and, thus, matches the version of the SHM using the grand canonical ensemble. 

The $^{3}$He/p and $^{3}$H/p ratios measured in p--Pb collisions, which cover the gap in the multiplicity between the existing measurements in pp and Pb--Pb collisions, favour a small correlation volume $V_\mathrm{c} = \text{d}V/\text{d}y$, while the ratios of the deuteron to the proton yield measured in pp collisions are more compatible with a larger correlation volume \cite{Vovchenko:2018fiy}.
The $^3$He/p ratio as a function of the mean charged-particle multiplicity density has a similar trend as the d/p ratio. However, the increase between the pp and the Pb--Pb results is about a factor of 3--4 larger for $^3$He/p than for d/p \cite{Acharya:2019rys}.
The simplified version of the CSM presented in this paper, which assumes a constant freeze-out temperature as a function of the system size, provides a good description of the multiplicity dependence of d/p, and of $^3$He/p, $^4$He/p and $^3_{\Lambda}$H/p in the high multiplicity range \cite{Vovchenko:2019kes}. However, this model shows some tensions with data for the $p/\pi$ and $K/\pi$ ratios and fails to describe the measured $\phi/\pi$ ratio \cite{Vovchenko:2019kes}.

With increasing mean charged-particle multiplicity density, the number of protons and neutrons produced in the collision also increases. The more protons and neutrons are available, the more likely nucleons can be close enough in phase space to form a nucleus. Therefore, an increasing trend for the $^3$He/p ratio as a function of the mean charged-particle multiplicity density is expected in the coalescence approach.
The measured ratio is compared to coalescence predictions \cite{Sun:2018mqq} which take the radii of the source and the emitted nucleus into account.
The case of three-body coalescence, where the nuclei are directly produced from protons and neutrons, as well as the expectation for two-body coalescence, where an intermediate formation of a deuteron is needed, are shown. For both coalescence approaches, the theoretical uncertainties are given by the uncertainty on the emission source radius.
Both calculations are consistent with the measurements within uncertainties while they are apparently below the experimental results for higher multiplicities. The measured $^3$He/p ratio shows a slight preference for the two-body coalescence approach, even though this is not yet conclusive due to the uncertainties on both the data and the theoretical description.

\subsection{Coalescence parameter ($B_3$)}

\begin{figure}[htb]
\centering
\includegraphics[width=0.7 \textwidth]{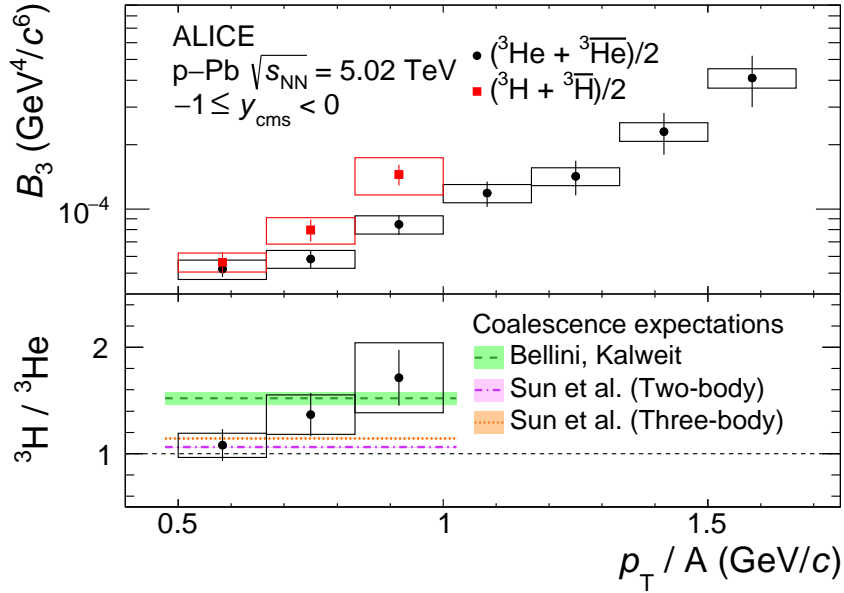}
\caption{Top: The coalescence parameter $B_3$ calculated using the average of $\mathrm{INEL}>0$ $^3$He and $^3\overline{\mathrm{He}}$ yields is shown together with the corresponding result for the average of the $^3$H and $^3\overline{\mathrm{H}}$ yields.
Bottom: The $^3$H/$^3$He yield ratio is shown together with the expectation values from three coalescence approaches \cite{Bellini:2018epz,Sun:2018mqq}. The uncertainties of the coalescence calculations, which are due to the theoretical uncertainties on the emission source radius, are denoted as shaded bands.
Statistical and systematic uncertainties are indicated by vertical bars and boxes, respectively.
}
\label{Figure:Coalencence}
\end{figure}

Following \autoref{eq:BA}, the coalescence parameter $B_3$ is obtained from the invariant yields of $^3$He or $^3$H and protons and is shown in \autoref{Figure:Coalencence} as a function of the transverse momentum per nucleon.

The ratio of the yields of (anti-)$^3$H and (anti-)$^3$He ($^3$H/$^3$He), which is shown in the bottom panel of \autoref{Figure:Coalencence}, is expected to be consistent with unity according to a naive coalescence approach. In more advanced coalescence calculations, that take into account the size of the emitting source and the nucleus, this ratio is expected to be above unity~\cite{Bellini:2018epz,Sun:2018mqq}. The difference in the coalescence expectations is mainly due to a different parametrization of the source radius as a function of the mean charged-particle multiplicity density. Another source of differences between the two coalescence approaches is the use of slightly different values for the radius of $^3$H.

The ratio is found to be in slightly better agreement with the coalescence expectations than with unity.
In the SHM approach the $^3$H/$^3$He ratio is expected to be consistent with unity. Thus, this observable is potentially useful not only to discriminate between different implementations of the coalescence approach but also with respect to SHMs. The increase of the $^3$H/$^3$He yield ratio with $\pt/A$ observed in data is not reflected in the theoretical predictions.

\begin{figure}[htb]
\centering
\includegraphics[width=0.7 \textwidth]{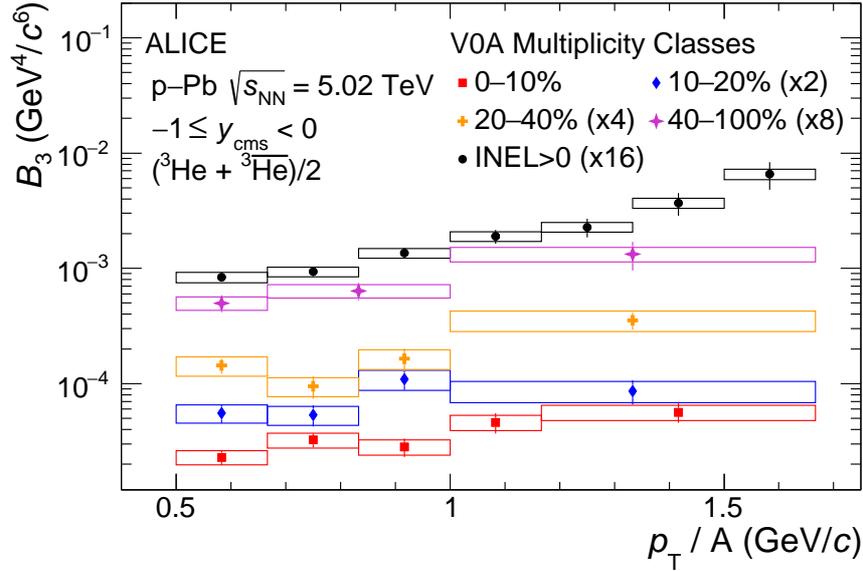}
\caption{The coalescence parameter $B_3$ calculated with the average of $^3$He and $^3\overline{\mathrm{He}}$ is shown for four multiplicity classes together with the $\mathrm{INEL}>0$ result. For better visibility, the distributions are scaled by different factors.
Statistical and systematic uncertainties are indicated by vertical bars and boxes, respectively.}
\label{Figure:CoalencenceVsMultiplicity}
\end{figure}

The coalescence parameter $B_3$ for $^{3}$He calculated for the four multiplicity classes analyzed is shown in \autoref{Figure:CoalencenceVsMultiplicity} as a function of the transverse momentum per nucleon.
A rising trend of $B_3$ with \pt/$A$ is observed in all multiplicity classes, contrary to the expectations of the naive coalescence approach, which predicts a constant $B_3$. 
This behavior can be partially understood as the effect coming from the change of the spectral shape of the protons with increasing mean charged-particle multiplicity density as explained for the measurement of deuterons in pp collisions $\sqrt{s} = 7 $ TeV \cite{Acharya:2019rgc}.
According to this, the coalescence parameter obtained in a wider charged-particle multiplicity interval exhibits an increasing trend with \pt/$A$ even though the coalescence parameter is flat in each smaller sub-interval.
To estimate the contribution of this effect to the $\mathrm{INEL}>0$ result, the coalescence parameter is recalculated considering the charged-particle multiplicity classes 0--5\%, 5--10\%, 10--20\%, 20--40\%, 40--60\%, 60--80\%, and 80--100\% as 
\begin{align}
   B_3^* = \frac{\sum_{i=0}^{n} (N_i/N) B^{i}_3 S^3_{\mathrm{p},i}} {\left(\sum_{i=0}^{n} (N_i/N) S_{\mathrm{p},i} \right)^3},
\end{align}
where $B^{i}_3$ and $S_{\mathrm{p},i} = 1/(2\pi\pt)\dndydpt$ are the coalescence parameter and the invariant yield of protons in the $i$th charged-particle multiplicity class, respectively.
The proton yields are taken from \cite{Abelev:2013haa}.
The weights $N_i/N$ are given by the fraction of events in the $i$th charged-particle multiplicity class.
The corresponding values of $B^{i}_3$ are assumed to be constant with \pt/$A$ and are obtained from the first \pt/$A$ interval of the corresponding measurements to ensure that $B_3^*$ starts from a value consistent with the one of the measurement.
The uncertainties on the re-calculated coalescence parameter $B_3^*$ are obtained by propagating the uncertainties on the proton spectra and on $B^{i}_3$. 
\begin{figure}[htb]
\centering
\includegraphics[width=0.7 \textwidth]{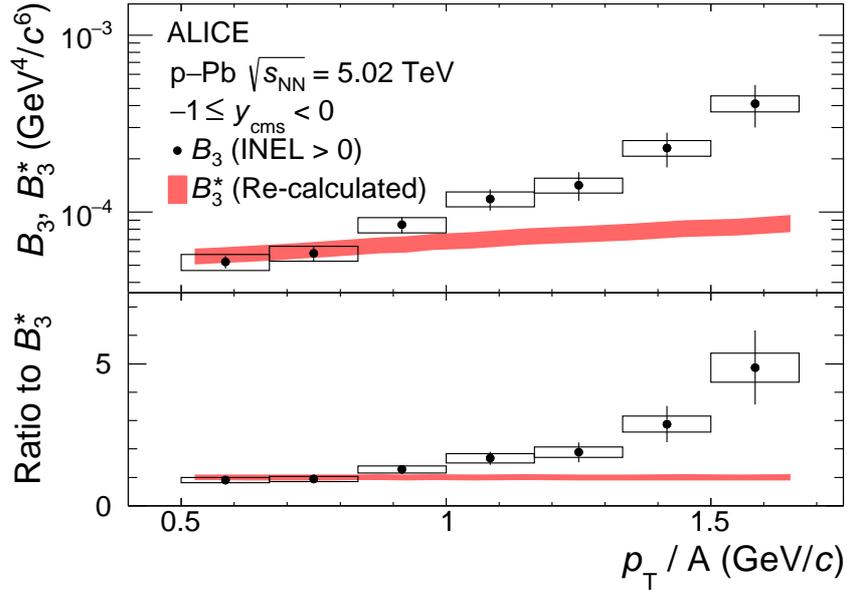}
\caption{Coalescence parameter $B_3$, obtained in $\mathrm{INEL}>0$ p--Pb collisions, compared to the recalculated coalescence parameter $B_3^*$ (see text for details).
Statistical and systematic uncertainties are indicated by vertical bars and boxes, respectively.}
\label{Figure:RecalculatedB3}
\end{figure}

As shown in \autoref{Figure:RecalculatedB3}, the resulting coalescence parameter $B_3^*$ increases by less than a factor of 2 while the measured $B_3$ increases by a factor of 8.
The data-driven procedure described above is based on the assumption that the coalescence parameter is constant with \pt/$A$ in the charged-particle multiplicity classes in which the protons are measured, which are up to 20$\%$ wide.
A stronger assumption is also tested, in which the $B^{i}_3$ are assumed to be \pt independent in multiplicity subintervals of $1\%$ width.
For this calculation, the proton spectra are obtained by interpolation of the measurements \cite{Abelev:2013haa}.
Both the data-driven approach and the interpolation method give consistent results.
This clearly indicates that a constant coalescence parameter with \pt/$A$ is too simplistic and cannot explain the observed increase in the measured $B_3$.

\begin{figure}[htb]
\centering
\includegraphics[width=0.49 \textwidth]{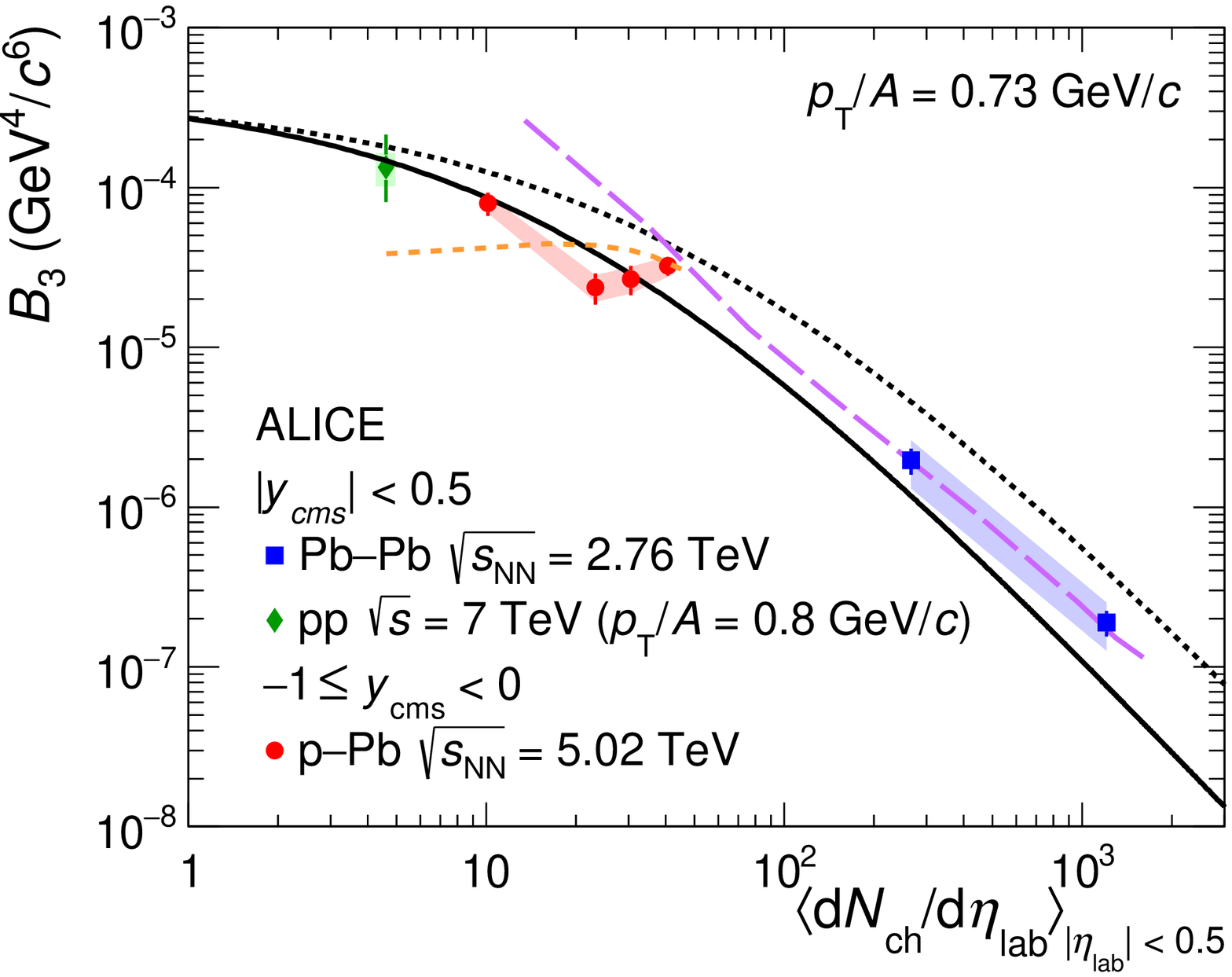}
\includegraphics[width=0.49 \textwidth]{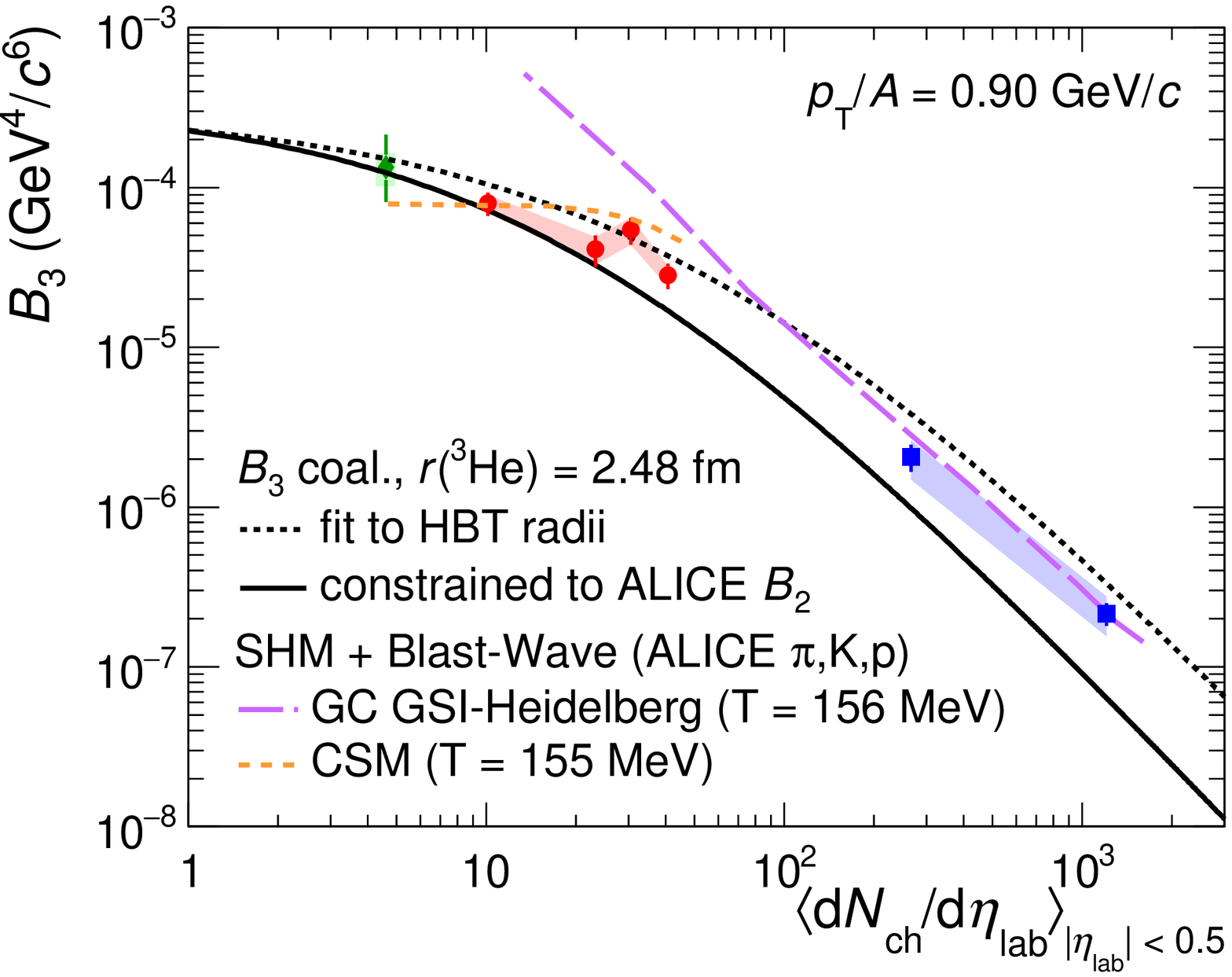}
\caption{The coalescence parameter $B_3$, calculated using the average of $^3$He and $^3\overline{\mathrm{He}}$, is shown as a function of the mean charged-particle multiplicity density for $\pt/A = 0.73$ GeV/$c$ (left) and $\pt/A = 0.90$ GeV/$c$ (right).
The coalescence parameter is shown with its statistical (line) and systematical (shaded area) uncertainties. In addition, the expectations from the coalescence and the SHM plus Blast-Wave approaches are shown \cite{Bellini:2018epz,Vovchenko:2018fiy}.}
\label{Figure:B3vsTheory}
\end{figure}

The multiplicity dependence of $B_3$ is compared to theoretical model calculations for $\pt/A = 0.73$ GeV/$c$ and $\pt/A = 0.90$ GeV/$c$ in \autoref{Figure:B3vsTheory}. The $B_3$ values for the measurements in pp, p--Pb, and Pb--Pb \cite{Acharya:2017fvb,Adam:2015vda} collisions are shown as a function of the mean charged-particle multiplicity density. In addition, the expected values for the coalescence approach taken from Ref.~\cite{Bellini:2018epz} are shown for two different parametrizations of the source radius as a function of the mean charged-particle multiplicity density. The two parametrizations can be understood as an indication of the validity band of the model description, which is expected to be more constrained with future measurements.
The measurements are compared to the expected values for the grand canonical version of the SHM, the GSI-Heidelberg model \cite{Andronic:2010qu,Andronic:2017pug}, assuming that the transverse momentum shape is given by a Blast-Wave parametrization obtained by a simultaneous fit to the pion, kaon, and proton spectra measured in Pb--Pb collisions \cite{Abelev:2013vea}. 
Since this model uses a grand-canonical description, it is applicable only for high mean charged-particle multiplicity densities. If canonical suppression is taken into account, the expected $B_3$ deviates from the grand canonical value, as indicated in \autoref{Figure:B3vsTheory} by exchanging the GSI-Heidelberg model with the CSM, Thermal-FIST \cite{Vovchenko:2018fiy}. The change to the canonical ensemble description extends the applicability of the model to intermediate mean charged-particle multiplicity densities. In the low mean charged-particle multiplicity density region, the assumption that the \pt shape of the nuclei follows the Blast-Wave parametrization breaks down. This is reflected by the larger deviation of the CSM plus Blast-Wave curve from the measured result in pp collisions for $\pt/A = 0.73$ GeV/$c$ compared to $\pt/A = 0.90$ GeV/$c$.

The best description of the coalescence parameter $B_3$ is given by the coalescence expectation for low mean charged-particle multiplicity densities and by the SHM for higher mean charged-particle multiplicity densities. The measurement of $B_3$ presented in this paper indicates a smooth transition between the regimes that are described by the two different approaches. The indication that the dominant production mechanism smoothly evolves with the charged-particle multiplicity density is consistent with previous ALICE measurements of the coalescence parameter $B_2$ in pp, p--Pb and Pb--Pb collisions at different center-of-mass energies \cite{Acharya:2019rgc,Acharya:2019rys,Adam:2015vda} and with the results on light (anti)nuclei elliptic flow in Pb--Pb collisions \cite{Acharya:2019ttn,Acharya:2017dmc}.

\subsection{Upper limit on the $^4\overline{\mathrm{He}}$ production}

An upper limit on the $^4\overline{\mathrm{He}}$ production in p--Pb collisions at \energy~=~5.02 TeV is estimated. The limit is based on the non-observation of $^4\overline{\mathrm{He}}$ candidates using the same track selection criteria as for $^3\overline{\mathrm{He}}$, except for the maximum distance-of-closest approach to the primary vertex. The DCA$_{\mathrm{xy}}$ is required to be smaller than 2.4 cm, while the DCA$_{\mathrm{z}}$ smaller than 3.2 cm.

The identification of $^4\overline{\mathrm{He}}$ is based on the time of flight, measured by the TOF detector, and the specific energy loss \dEdx in the TPC. These measurements are required to be within $\pm 5 \sigma_\mathrm{TOF}$ and $\pm 3 \sigma_\mathrm{TPC}$ from the expected values. 
The analysis is performed in the transverse momentum interval $2 \leq \pt < 10$ GeV/$c$. 

\begin{figure}[htb]
\centering
\includegraphics[width=0.49 \textwidth]{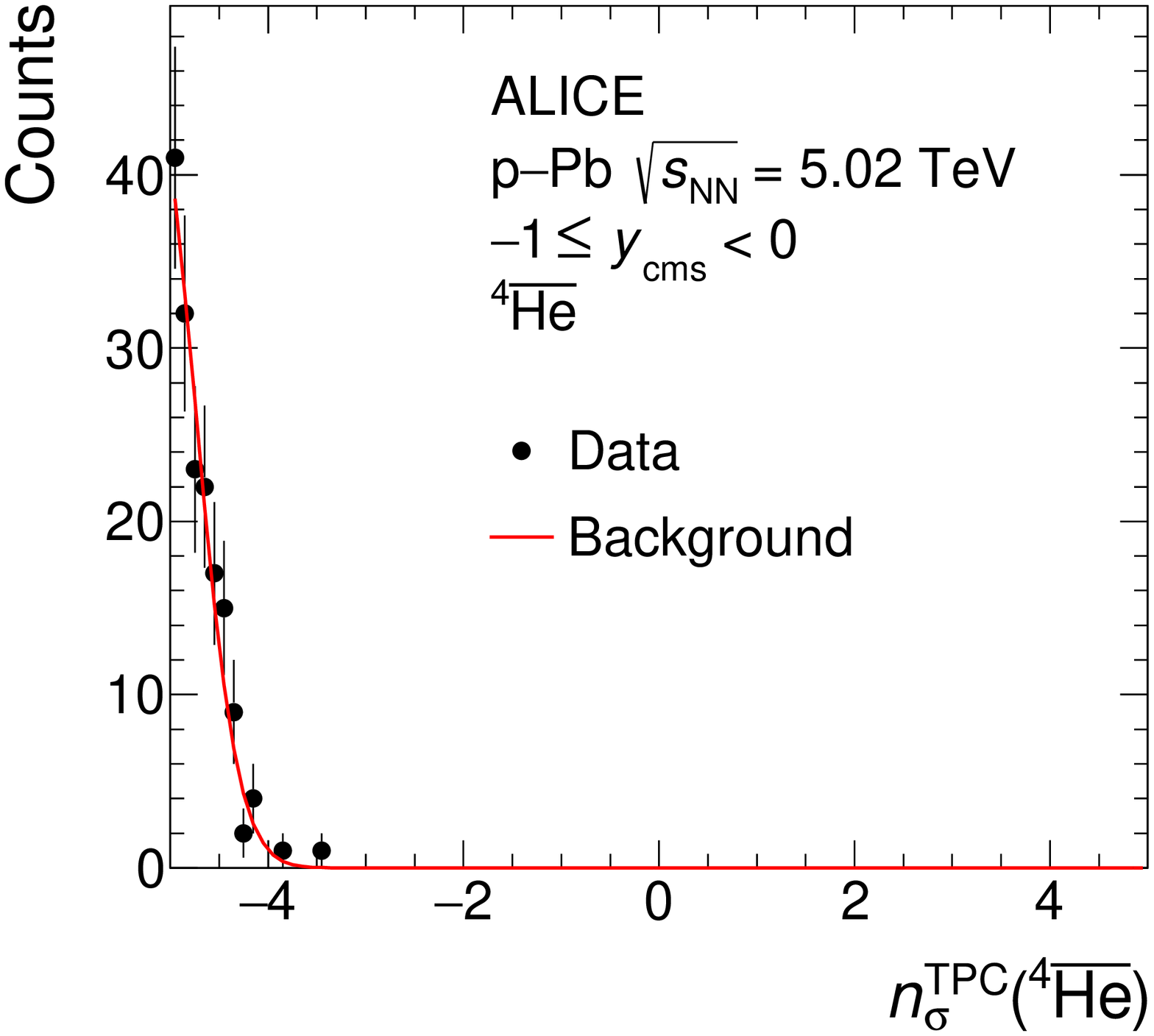}
\includegraphics[width=0.49 \textwidth]{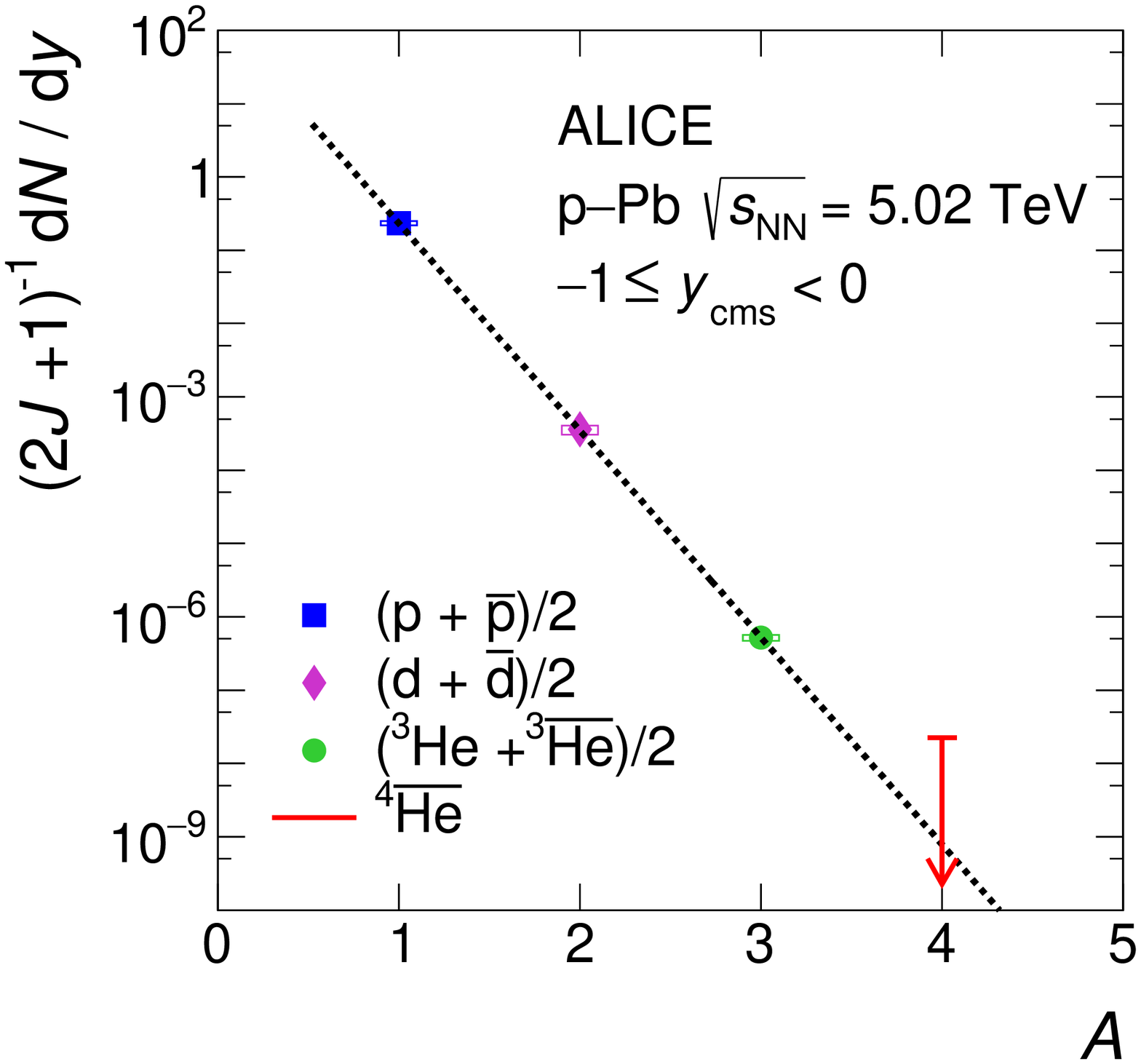}
\caption{Left: The distribution of the specific ionization energy loss (\dEdx) in the TPC for candidate tracks compared to the expected one for $^4\overline{\text{He}}$, $n_{\sigma}^\mathrm{TPC}$, in the \pt range from 2 to 10 GeV/$c$. The background is fitted with a Gaussian function (red line).
Right: The production yield \dNdy corrected for the spin degeneracy factor $(2J +1)$ as a function of the mass number for minimum-bias p--Pb collisions at \energy~=~5.02 TeV is shown. The empty boxes represent the total systematic uncertainty while the statistical uncertainties are shown by the vertical bars. The line represents a fit with an exponential function.
}
\label{Figure:UpperLimit}
\end{figure}

The left panel of \autoref{Figure:UpperLimit} shows the distribution of the specific energy loss compared to the expected one for $^4\mathrm{He}$ ($n_\sigma^\mathrm{TPC}$) after the preselection using the TOF.
The distribution at $n_\sigma^\mathrm{TPC} < -3$, corresponding to $^3\overline{\mathrm{He}}$ candidates, is fitted with a Gaussian function and extrapolated to the signal region, defined by the range $[-3,3]$. The expected background in the signal region is $ 1 \times 10^{-5}$. 
The expected background and the nonobservation of candidates in the signal region are used to calculate the upper limit at 90$\%$ confidence level using the Feldmann-Cousins approach \cite{Feldman:1997qc}. 
The resulting number is corrected for the product of the acceptance and the reconstruction efficiency, the rapidity range, and the number of events selected.
The product of the acceptance and the reconstruction efficiency was obtained
as the average of the values found in smaller \pt intervals weighted with the expected shape of the spectrum of $^4$He. For the latter, a \mt-exponential parametrization of the $^4\overline{\mathrm{He}}$ \pt spectrum, with parameters identical to those of $^{3}$He except for the mass, which is set equal to the $^{4}$He mass, was used.
The obtained value for the upper limit is extrapolated to the full \pt range using the \mt-exponential parametrization of the $^4\overline{\mathrm{He}}$ \pt spectrum. A systematic uncertainty of 20\%, similar to that of the measurement in Pb--Pb collisions at \energy = 2.76 TeV \cite{Acharya:2017bso}, is taken into account following the procedure described in Refs.~\cite{Conrad:2002kn, Hill:2003jk}. 
 
The upper limit on the $^4\overline{\mathrm{He}}$ total yield (\dNdy) in p--Pb collisions at \energy~=~5.02~TeV is found to be $2.3 \times 10^{-8}$ at 90\% confidence level. The upper limit is shown in the right panel of \autoref{Figure:UpperLimit} together with the measured \dNdy corrected for the spin degeneracy factor, $2J +1$, of (anti)protons \cite{Abelev:2013haa}, \mbox{(anti-)}deuterons \cite{Acharya:2019rys}, and (anti-)$^{3}$He.
The upper limit is compatible with a penalty factor, i.e. the suppression of the yield for each additional nucleon, of $668 \pm 45$ obtained by fitting the measurements of the proton, deuteron, and $^{3}$He yields with an exponential function. The value of the penalty factor is consistent with the one obtained in previous measurements \cite{Acharya:2019rys}.
Taking into account this penalty factor, the expected yield of (anti-)$^{4}$He is about $8 \times 10^{-10}$.

\section{Conclusions}
The \pt-differential yields of $^3$H and $^3$He nuclei and their anti-nuclei were measured in p--Pb collisions at \energy~= 5.02 TeV. For (anti-)$^3$He, the production was studied in different classes of mean charged-particle multiplicity density.

A consistent comparison between experimental results and the canonical statistical hadronization model as well as the coalescence calculations was done for the same observable.
The $^3$He/p ratio measured in p--Pb collisions at \energy~= 5.02 TeV bridges the gap between existing measurements in pp and Pb--Pb collisions and is overall in good agreement with the theoretical descriptions.
Despite the agreement of the measurement and the CSM model, there is some tension due to the bad matching of the predicted and measured $p/\pi$ and $K/\pi$ ratios and the failure to describe the measured $\phi/\pi$ ratio.
The coalescence approach has some difficulties to describe the measurements at high mean charged-particle multiplicity densities.

The coalescence parameter is measured as a function of the transverse momentum per nucleon. The result presented in this paper shows an increasing trend which is not compatible with naive coalescence approaches and cannot be explained by the hardening of the proton spectra with charged-particle multiplicity \cite{Acharya:2019rgc}.
The presented results clearly imply an increasing trend of $B_3$ also in small charged-particle multiplicity intervals.
The increasing coalescence probability with increasing \pt implies stronger correlation of nucleons in momentum and space at high \pt.
A clear answer to the question about the origin of the increasing trend with \pt/$A$ observed for the different multiplicity classes requires a larger data set and more detailed theoretical descriptions.

The coalescence parameter $B_3$ is also measured as a function of the mean charged-particle multiplicity density and compared to expectations from the grand canonical and the canonical versions of the SHM. The use of the Blast-Wave parametrizations to define the \pt shape breaks down for low multiplicities, which leads to larger discrepancies between the CSM and the measurements for the two $\pt/A$ intervals shown.
In addition, the measurements are compared to the coalescence expectations for two different parametrizations of the source radius as a function of the mean charged-particle multiplicity density.
The presented measurements are in agreement with the coalescence description as well as the SHM description within theoretical and experimental uncertainties.
The data indicate a smooth transition between the regimes described best by the coalescence approach and the Statistical Hadronization Model.

These measurements provide the possibility to test the dependence of the production rate on the nuclear radius by direct comparison of isospin partner nuclei in the same data set at the LHC.
The $^3$H/$^3$He ratio is sensitive to the production mechanism within the coalescence approach.
Even though the measurement presented in this paper is not yet conclusive due to the large uncertainties, an apparent deviation from unity can be observed which would slightly favour the coalescence description including the dependence on the radii of the nucleus and the emitting source.

An upper limit on the total production yield of $^4\overline{\mathrm{He}}$ in p--Pb collisions at \energy~= 5.02 TeV was found which is about two orders of magnitude above the expected result obtained from the exponential fit of the proton, deuteron, and $^3$He yields.

\newenvironment{acknowledgement}{\relax}{\relax}
\begin{acknowledgement}
\section*{Acknowledgements}

The ALICE Collaboration would like to thank all its engineers and technicians for their invaluable contributions to the construction of the experiment and the CERN accelerator teams for the outstanding performance of the LHC complex.
The ALICE Collaboration gratefully acknowledges the resources and support provided by all Grid centres and the Worldwide LHC Computing Grid (WLCG) collaboration.
The ALICE Collaboration acknowledges the following funding agencies for their support in building and running the ALICE detector:
A. I. Alikhanyan National Science Laboratory (Yerevan Physics Institute) Foundation (ANSL), State Committee of Science and World Federation of Scientists (WFS), Armenia;
Austrian Academy of Sciences, Austrian Science Fund (FWF): [M 2467-N36] and Nationalstiftung f\"{u}r Forschung, Technologie und Entwicklung, Austria;
Ministry of Communications and High Technologies, National Nuclear Research Center, Azerbaijan;
Conselho Nacional de Desenvolvimento Cient\'{\i}fico e Tecnol\'{o}gico (CNPq), Financiadora de Estudos e Projetos (Finep), Funda\c{c}\~{a}o de Amparo \`{a} Pesquisa do Estado de S\~{a}o Paulo (FAPESP) and Universidade Federal do Rio Grande do Sul (UFRGS), Brazil;
Ministry of Education of China (MOEC) , Ministry of Science \& Technology of China (MSTC) and National Natural Science Foundation of China (NSFC), China;
Ministry of Science and Education and Croatian Science Foundation, Croatia;
Centro de Aplicaciones Tecnol\'{o}gicas y Desarrollo Nuclear (CEADEN), Cubaenerg\'{\i}a, Cuba;
Ministry of Education, Youth and Sports of the Czech Republic, Czech Republic;
The Danish Council for Independent Research | Natural Sciences, the VILLUM FONDEN and Danish National Research Foundation (DNRF), Denmark;
Helsinki Institute of Physics (HIP), Finland;
Commissariat \`{a} l'Energie Atomique (CEA), Institut National de Physique Nucl\'{e}aire et de Physique des Particules (IN2P3) and Centre National de la Recherche Scientifique (CNRS) and R\'{e}gion des  Pays de la Loire, France;
Bundesministerium f\"{u}r Bildung und Forschung (BMBF) and GSI Helmholtzzentrum f\"{u}r Schwerionenforschung GmbH, Germany;
General Secretariat for Research and Technology, Ministry of Education, Research and Religions, Greece;
National Research, Development and Innovation Office, Hungary;
Department of Atomic Energy Government of India (DAE), Department of Science and Technology, Government of India (DST), University Grants Commission, Government of India (UGC) and Council of Scientific and Industrial Research (CSIR), India;
Indonesian Institute of Science, Indonesia;
Centro Fermi - Museo Storico della Fisica e Centro Studi e Ricerche Enrico Fermi and Istituto Nazionale di Fisica Nucleare (INFN), Italy;
Institute for Innovative Science and Technology , Nagasaki Institute of Applied Science (IIST), Japanese Ministry of Education, Culture, Sports, Science and Technology (MEXT) and Japan Society for the Promotion of Science (JSPS) KAKENHI, Japan;
Consejo Nacional de Ciencia (CONACYT) y Tecnolog\'{i}a, through Fondo de Cooperaci\'{o}n Internacional en Ciencia y Tecnolog\'{i}a (FONCICYT) and Direcci\'{o}n General de Asuntos del Personal Academico (DGAPA), Mexico;
Nederlandse Organisatie voor Wetenschappelijk Onderzoek (NWO), Netherlands;
The Research Council of Norway, Norway;
Commission on Science and Technology for Sustainable Development in the South (COMSATS), Pakistan;
Pontificia Universidad Cat\'{o}lica del Per\'{u}, Peru;
Ministry of Science and Higher Education and National Science Centre, Poland;
Korea Institute of Science and Technology Information and National Research Foundation of Korea (NRF), Republic of Korea;
Ministry of Education and Scientific Research, Institute of Atomic Physics and Ministry of Research and Innovation and Institute of Atomic Physics, Romania;
Joint Institute for Nuclear Research (JINR), Ministry of Education and Science of the Russian Federation, National Research Centre Kurchatov Institute, Russian Science Foundation and Russian Foundation for Basic Research, Russia;
Ministry of Education, Science, Research and Sport of the Slovak Republic, Slovakia;
National Research Foundation of South Africa, South Africa;
Swedish Research Council (VR) and Knut \& Alice Wallenberg Foundation (KAW), Sweden;
European Organization for Nuclear Research, Switzerland;
Suranaree University of Technology (SUT), National Science and Technology Development Agency (NSDTA) and Office of the Higher Education Commission under NRU project of Thailand, Thailand;
Turkish Atomic Energy Agency (TAEK), Turkey;
National Academy of  Sciences of Ukraine, Ukraine;
Science and Technology Facilities Council (STFC), United Kingdom;
National Science Foundation of the United States of America (NSF) and United States Department of Energy, Office of Nuclear Physics (DOE NP), United States of America.    
\end{acknowledgement}

\bibliographystyle{utphys}   
\bibliography{NucleiBiblio}

\newpage
\appendix
\section{The ALICE Collaboration}
\label{app:collab}

\begingroup
\small
\begin{flushleft}
S.~Acharya$^{\rm 141}$, 
D.~Adamov\'{a}$^{\rm 94}$, 
A.~Adler$^{\rm 74}$, 
J.~Adolfsson$^{\rm 80}$, 
M.M.~Aggarwal$^{\rm 99}$, 
G.~Aglieri Rinella$^{\rm 33}$, 
M.~Agnello$^{\rm 30}$, 
N.~Agrawal$^{\rm 10,53}$, 
Z.~Ahammed$^{\rm 141}$, 
S.~Ahmad$^{\rm 16}$
% ...
\renewcommand\labelenumi{\textsuperscript{\theenumi}~}

\section*{Affiliation notes}
\renewcommand\theenumi{\roman{enumi}}
\begin{Authlist}
\item \Adef{org*}Deceased
\item \Adef{orgI}Dipartimento DET del Politecnico di Torino, Turin, Italy
\item \Adef{orgII}M.V. Lomonosov Moscow State University, D.V. Skobeltsyn Institute of Nuclear, Physics, Moscow, Russia
\item \Adef{orgIII}Department of Applied Physics, Aligarh Muslim University, Aligarh, India
\item \Adef{orgIV}Institute of Theoretical Physics, University of Wroclaw, Poland
\end{Authlist}

\section*{Collaboration Institutes}
\renewcommand\theenumi{\arabic{enumi}~}
\begin{Authlist}
\item \Idef{org1}A.I. Alikhanyan National Science Laboratory (Yerevan Physics Institute) Foundation, Yerevan, Armenia
\item \Idef{org2}Bogolyubov Institute for Theoretical Physics, National Academy of Sciences of Ukraine, Kiev, Ukraine
\item \Idef{org3}Bose Institute, Department of Physics  and Centre for Astroparticle Physics and Space Science (CAPSS), Kolkata, India
\item \Idef{org4}Budker Institute for Nuclear Physics, Novosibirsk, Russia
\item \Idef{org5}California Polytechnic State University, San Luis Obispo, California, United States
\item \Idef{org6}Central China Normal University, Wuhan, China
\item \Idef{org7}Centre de Calcul de l'IN2P3, Villeurbanne, Lyon, France
\item \Idef{org8}Centro de Aplicaciones Tecnol\'{o}gicas y Desarrollo Nuclear (CEADEN), Havana, Cuba
\item \Idef{org9}Centro de Investigaci\'{o}n y de Estudios Avanzados (CINVESTAV), Mexico City and M\'{e}rida, Mexico
\item \Idef{org10}Centro Fermi - Museo Storico della Fisica e Centro Studi e Ricerche ``Enrico Fermi', Rome, Italy
\item \Idef{org11}Chicago State University, Chicago, Illinois, United States
\item \Idef{org12}China Institute of Atomic Energy, Beijing, China
\item \Idef{org13}Comenius University Bratislava, Faculty of Mathematics, Physics and Informatics, Bratislava, Slovakia
\item \Idef{org14}COMSATS University Islamabad, Islamabad, Pakistan
\item \Idef{org15}Creighton University, Omaha, Nebraska, United States
\item \Idef{org16}Department of Physics, Aligarh Muslim University, Aligarh, India
\item \Idef{org17}Department of Physics, Pusan National University, Pusan, Republic of Korea
\item \Idef{org18}Department of Physics, Sejong University, Seoul, Republic of Korea
\item \Idef{org19}Department of Physics, University of California, Berkeley, California, United States
\item \Idef{org20}Department of Physics, University of Oslo, Oslo, Norway
\item \Idef{org21}Department of Physics and Technology, University of Bergen, Bergen, Norway
\item \Idef{org22}Dipartimento di Fisica dell'Universit\`{a} 'La Sapienza' and Sezione INFN, Rome, Italy
\item \Idef{org23}Dipartimento di Fisica dell'Universit\`{a} and Sezione INFN, Cagliari, Italy
\item \Idef{org24}Dipartimento di Fisica dell'Universit\`{a} and Sezione INFN, Trieste, Italy
\item \Idef{org25}Dipartimento di Fisica dell'Universit\`{a} and Sezione INFN, Turin, Italy
\item \Idef{org26}Dipartimento di Fisica e Astronomia dell'Universit\`{a} and Sezione INFN, Bologna, Italy
\item \Idef{org27}Dipartimento di Fisica e Astronomia dell'Universit\`{a} and Sezione INFN, Catania, Italy
\item \Idef{org28}Dipartimento di Fisica e Astronomia dell'Universit\`{a} and Sezione INFN, Padova, Italy
\item \Idef{org29}Dipartimento di Fisica `E.R.~Caianiello' dell'Universit\`{a} and Gruppo Collegato INFN, Salerno, Italy
\item \Idef{org30}Dipartimento DISAT del Politecnico and Sezione INFN, Turin, Italy
\item \Idef{org31}Dipartimento di Scienze e Innovazione Tecnologica dell'Universit\`{a} del Piemonte Orientale and INFN Sezione di Torino, Alessandria, Italy
\item \Idef{org32}Dipartimento Interateneo di Fisica `M.~Merlin' and Sezione INFN, Bari, Italy
\item \Idef{org33}European Organization for Nuclear Research (CERN), Geneva, Switzerland
\item \Idef{org34}Faculty of Electrical Engineering, Mechanical Engineering and Naval Architecture, University of Split, Split, Croatia
\item \Idef{org35}Faculty of Engineering and Science, Western Norway University of Applied Sciences, Bergen, Norway
\item \Idef{org36}Faculty of Nuclear Sciences and Physical Engineering, Czech Technical University in Prague, Prague, Czech Republic
\item \Idef{org37}Faculty of Science, P.J.~\v{S}af\'{a}rik University, Ko\v{s}ice, Slovakia
\item \Idef{org38}Frankfurt Institute for Advanced Studies, Johann Wolfgang Goethe-Universit\"{a}t Frankfurt, Frankfurt, Germany
\item \Idef{org39}Fudan University, Shanghai, China
\item \Idef{org40}Gangneung-Wonju National University, Gangneung, Republic of Korea
\item \Idef{org41}Gauhati University, Department of Physics, Guwahati, India
\item \Idef{org42}Helmholtz-Institut f\"{u}r Strahlen- und Kernphysik, Rheinische Friedrich-Wilhelms-Universit\"{a}t Bonn, Bonn, Germany
\item \Idef{org43}Helsinki Institute of Physics (HIP), Helsinki, Finland
\item \Idef{org44}High Energy Physics Group,  Universidad Aut\'{o}noma de Puebla, Puebla, Mexico
\item \Idef{org45}Hiroshima University, Hiroshima, Japan
\item \Idef{org46}Hochschule Worms, Zentrum  f\"{u}r Technologietransfer und Telekommunikation (ZTT), Worms, Germany
\item \Idef{org47}Horia Hulubei National Institute of Physics and Nuclear Engineering, Bucharest, Romania
\item \Idef{org48}Indian Institute of Technology Bombay (IIT), Mumbai, India
\item \Idef{org49}Indian Institute of Technology Indore, Indore, India
\item \Idef{org50}Indonesian Institute of Sciences, Jakarta, Indonesia
\item \Idef{org51}INFN, Laboratori Nazionali di Frascati, Frascati, Italy
\item \Idef{org52}INFN, Sezione di Bari, Bari, Italy
\item \Idef{org53}INFN, Sezione di Bologna, Bologna, Italy
\item \Idef{org54}INFN, Sezione di Cagliari, Cagliari, Italy
\item \Idef{org55}INFN, Sezione di Catania, Catania, Italy
\item \Idef{org56}INFN, Sezione di Padova, Padova, Italy
\item \Idef{org57}INFN, Sezione di Roma, Rome, Italy
\item \Idef{org58}INFN, Sezione di Torino, Turin, Italy
\item \Idef{org59}INFN, Sezione di Trieste, Trieste, Italy
\item \Idef{org60}Inha University, Incheon, Republic of Korea
\item \Idef{org61}Institut de Physique Nucl\'{e}aire d'Orsay (IPNO), Institut National de Physique Nucl\'{e}aire et de Physique des Particules (IN2P3/CNRS), Universit\'{e} de Paris-Sud, Universit\'{e} Paris-Saclay, Orsay, France
\item \Idef{org62}Institute for Nuclear Research, Academy of Sciences, Moscow, Russia
\item \Idef{org63}Institute for Subatomic Physics, Utrecht University/Nikhef, Utrecht, Netherlands
\item \Idef{org64}Institute of Experimental Physics, Slovak Academy of Sciences, Ko\v{s}ice, Slovakia
\item \Idef{org65}Institute of Physics, Homi Bhabha National Institute, Bhubaneswar, India
\item \Idef{org66}Institute of Physics of the Czech Academy of Sciences, Prague, Czech Republic
\item \Idef{org67}Institute of Space Science (ISS), Bucharest, Romania
\item \Idef{org68}Institut f\"{u}r Kernphysik, Johann Wolfgang Goethe-Universit\"{a}t Frankfurt, Frankfurt, Germany
\item \Idef{org69}Instituto de Ciencias Nucleares, Universidad Nacional Aut\'{o}noma de M\'{e}xico, Mexico City, Mexico
\item \Idef{org70}Instituto de F\'{i}sica, Universidade Federal do Rio Grande do Sul (UFRGS), Porto Alegre, Brazil
\item \Idef{org71}Instituto de F\'{\i}sica, Universidad Nacional Aut\'{o}noma de M\'{e}xico, Mexico City, Mexico
\item \Idef{org72}iThemba LABS, National Research Foundation, Somerset West, South Africa
\item \Idef{org73}Jeonbuk National University, Jeonju, Republic of Korea
\item \Idef{org74}Johann-Wolfgang-Goethe Universit\"{a}t Frankfurt Institut f\"{u}r Informatik, Fachbereich Informatik und Mathematik, Frankfurt, Germany
\item \Idef{org75}Joint Institute for Nuclear Research (JINR), Dubna, Russia
\item \Idef{org76}Korea Institute of Science and Technology Information, Daejeon, Republic of Korea
\item \Idef{org77}KTO Karatay University, Konya, Turkey
\item \Idef{org78}Laboratoire de Physique Subatomique et de Cosmologie, Universit\'{e} Grenoble-Alpes, CNRS-IN2P3, Grenoble, France
\item \Idef{org79}Lawrence Berkeley National Laboratory, Berkeley, California, United States
\item \Idef{org80}Lund University Department of Physics, Division of Particle Physics, Lund, Sweden
\item \Idef{org81}Nagasaki Institute of Applied Science, Nagasaki, Japan
\item \Idef{org82}Nara Women{'}s University (NWU), Nara, Japan
\item \Idef{org83}National and Kapodistrian University of Athens, School of Science, Department of Physics , Athens, Greece
\item \Idef{org84}National Centre for Nuclear Research, Warsaw, Poland
\item \Idef{org85}National Institute of Science Education and Research, Homi Bhabha National Institute, Jatni, India
\item \Idef{org86}National Nuclear Research Center, Baku, Azerbaijan
\item \Idef{org87}National Research Centre Kurchatov Institute, Moscow, Russia
\item \Idef{org88}Niels Bohr Institute, University of Copenhagen, Copenhagen, Denmark
\item \Idef{org89}Nikhef, National institute for subatomic physics, Amsterdam, Netherlands
\item \Idef{org90}NRC Kurchatov Institute IHEP, Protvino, Russia
\item \Idef{org91}NRC «Kurchatov Institute»  - ITEP, Moscow, Russia
\item \Idef{org92}NRNU Moscow Engineering Physics Institute, Moscow, Russia
\item \Idef{org93}Nuclear Physics Group, STFC Daresbury Laboratory, Daresbury, United Kingdom
\item \Idef{org94}Nuclear Physics Institute of the Czech Academy of Sciences, \v{R}e\v{z} u Prahy, Czech Republic
\item \Idef{org95}Oak Ridge National Laboratory, Oak Ridge, Tennessee, United States
\item \Idef{org96}Ohio State University, Columbus, Ohio, United States
\item \Idef{org97}Petersburg Nuclear Physics Institute, Gatchina, Russia
\item \Idef{org98}Physics department, Faculty of science, University of Zagreb, Zagreb, Croatia
\item \Idef{org99}Physics Department, Panjab University, Chandigarh, India
\item \Idef{org100}Physics Department, University of Jammu, Jammu, India
\item \Idef{org101}Physics Department, University of Rajasthan, Jaipur, India
\item \Idef{org102}Physikalisches Institut, Eberhard-Karls-Universit\"{a}t T\"{u}bingen, T\"{u}bingen, Germany
\item \Idef{org103}Physikalisches Institut, Ruprecht-Karls-Universit\"{a}t Heidelberg, Heidelberg, Germany
\item \Idef{org104}Physik Department, Technische Universit\"{a}t M\"{u}nchen, Munich, Germany
\item \Idef{org105}Politecnico di Bari, Bari, Italy
\item \Idef{org106}Research Division and ExtreMe Matter Institute EMMI, GSI Helmholtzzentrum f\"ur Schwerionenforschung GmbH, Darmstadt, Germany
\item \Idef{org107}Rudjer Bo\v{s}kovi\'{c} Institute, Zagreb, Croatia
\item \Idef{org108}Russian Federal Nuclear Center (VNIIEF), Sarov, Russia
\item \Idef{org109}Saha Institute of Nuclear Physics, Homi Bhabha National Institute, Kolkata, India
\item \Idef{org110}School of Physics and Astronomy, University of Birmingham, Birmingham, United Kingdom
\item \Idef{org111}Secci\'{o}n F\'{\i}sica, Departamento de Ciencias, Pontificia Universidad Cat\'{o}lica del Per\'{u}, Lima, Peru
\item \Idef{org112}St. Petersburg State University, St. Petersburg, Russia
\item \Idef{org113}Stefan Meyer Institut f\"{u}r Subatomare Physik (SMI), Vienna, Austria
\item \Idef{org114}SUBATECH, IMT Atlantique, Universit\'{e} de Nantes, CNRS-IN2P3, Nantes, France
\item \Idef{org115}Suranaree University of Technology, Nakhon Ratchasima, Thailand
\item \Idef{org116}Technical University of Ko\v{s}ice, Ko\v{s}ice, Slovakia
\item \Idef{org117}Technische Universit\"{a}t M\"{u}nchen, Excellence Cluster 'Universe', Munich, Germany
\item \Idef{org118}The Henryk Niewodniczanski Institute of Nuclear Physics, Polish Academy of Sciences, Cracow, Poland
\item \Idef{org119}The University of Texas at Austin, Austin, Texas, United States
\item \Idef{org120}Universidad Aut\'{o}noma de Sinaloa, Culiac\'{a}n, Mexico
\item \Idef{org121}Universidade de S\~{a}o Paulo (USP), S\~{a}o Paulo, Brazil
\item \Idef{org122}Universidade Estadual de Campinas (UNICAMP), Campinas, Brazil
\item \Idef{org123}Universidade Federal do ABC, Santo Andre, Brazil
\item \Idef{org124}University of Cape Town, Cape Town, South Africa
\item \Idef{org125}University of Houston, Houston, Texas, United States
\item \Idef{org126}University of Jyv\"{a}skyl\"{a}, Jyv\"{a}skyl\"{a}, Finland
\item \Idef{org127}University of Liverpool, Liverpool, United Kingdom
\item \Idef{org128}University of Science and Techonology of China, Hefei, China
\item \Idef{org129}University of South-Eastern Norway, Tonsberg, Norway
\item \Idef{org130}University of Tennessee, Knoxville, Tennessee, United States
\item \Idef{org131}University of the Witwatersrand, Johannesburg, South Africa
\item \Idef{org132}University of Tokyo, Tokyo, Japan
\item \Idef{org133}University of Tsukuba, Tsukuba, Japan
\item \Idef{org134}Universit\'{e} Clermont Auvergne, CNRS/IN2P3, LPC, Clermont-Ferrand, France
\item \Idef{org135}Universit\'{e} de Lyon, Universit\'{e} Lyon 1, CNRS/IN2P3, IPN-Lyon, Villeurbanne, Lyon, France
\item \Idef{org136}Universit\'{e} de Strasbourg, CNRS, IPHC UMR 7178, F-67000 Strasbourg, France, Strasbourg, France
\item \Idef{org137}Universit\'{e} Paris-Saclay Centre d'Etudes de Saclay (CEA), IRFU, D\'{e}partment de Physique Nucl\'{e}aire (DPhN), Saclay, France
\item \Idef{org138}Universit\`{a} degli Studi di Foggia, Foggia, Italy
\item \Idef{org139}Universit\`{a} degli Studi di Pavia, Pavia, Italy
\item \Idef{org140}Universit\`{a} di Brescia, Brescia, Italy
\item \Idef{org141}Variable Energy Cyclotron Centre, Homi Bhabha National Institute, Kolkata, India
\item \Idef{org142}Warsaw University of Technology, Warsaw, Poland
\item \Idef{org143}Wayne State University, Detroit, Michigan, United States
\item \Idef{org144}Westf\"{a}lische Wilhelms-Universit\"{a}t M\"{u}nster, Institut f\"{u}r Kernphysik, M\"{u}nster, Germany
\item \Idef{org145}Wigner Research Centre for Physics, Budapest, Hungary
\item \Idef{org146}Yale University, New Haven, Connecticut, United States
\item \Idef{org147}Yonsei University, Seoul, Republic of Korea
\end{Authlist}
\endgroup  
\end{document}